\documentclass[]{scrartcl}
\pdfoutput=1 

\usepackage[utf8]{inputenc}
\usepackage{amsmath}
\usepackage{amssymb}
\usepackage{natbib}
\usepackage{bbm}
\usepackage{color}
\usepackage{amsthm}
\usepackage{graphicx}
\usepackage{bm}
\usepackage{subfig}
\usepackage{url}
\usepackage{graphicx}

\usepackage{authblk}

\usepackage{amsfonts}
\usepackage{xcolor}
\definecolor{C0}{HTML}{3182ce}
\usepackage[breaklinks=true,colorlinks,bookmarks=false,citecolor=C0]{hyperref}
\usepackage[margin=1in]{geometry}
\usepackage{setspace}
\usepackage{algorithm}
\usepackage{algpseudocode}

\newtheorem{theorem}{Theorem}[section]

\newtheorem{lemma}[theorem]{Lemma}
\newtheorem{assumption}{Assumption}

\newtheorem{remark}{Remark}
\newtheorem{example}{Example}

\def\1{\mathbbm{1}}

\usepackage{xcolor}

\graphicspath{{}{figures/}}

\newcommand{\norm}[1]{\left\lVert#1\right\rVert}

\title{Asymptotics of approximate Bayesian computation when summary statistics converge at heterogeneous rates}

\author[1]{Caroline Lawless}
\author[2,3]{Christian P. Robert}
\author[1,2]{Judith Rousseau}
\author[2]{Robin J. Ryder}

\affil[1]{Department of Statistics, University of Oxford}
\affil[2]{CEREMADE, Université Paris-Dauphine PSL}
\affil[3]{Department of Statistics, University of Warwick}

\begin{document}

\maketitle

\section{Introduction} \label{intro}

\textit{Likelihood-free} methods in Bayesian statistics are methods for posterior inference in situations where likelihoods are intractable or unavailable in closed form, but may be simulated from. Such situations are typical of real-life applications, when models are defined by complex generative processes. This can result in likelihoods involving high dimensional integrals which are impossible to compute in a reasonable amount of time. 
Examples of such likelihood-free methods include simulated methods of moments \citep{duffie1990simulated}, indirect inference \citep{gourieroux1993indirect}, synthetic likelihood \citep{wood2010statistical} and approximate Bayesian computation (ABC) \citep{sisson2018handbook}. In this work, we focus on the latter. At its core, ABC relies on simulating many data sets from the prior predictive. The data sets are summarized by a low dimensional statistic, and only those within a small pseudo distance (the tolerance) of the observed data are kept.

ABC was first introduced in the context of population genetics \citep{pritchard1999population}. It has since then  been applied in research areas as diverse as population genetics \citep{pritchard1999population}, protein networks \citep{ratmann2007using}, epidemiology \citep{tanaka2006using} , inference for extremes \citep{bortot2007inference}, dynamical systems \citep{toni2009approximate}, and Gibbs random fields \citep{grelaud2009abc}. Due to its increasing popularity in applied statistics, recent research has focused on the theoretical properties of ABC methods.

\cite{fearnhead2012constructing}  consider the question of summary statistic choice, and find that summary statistics should ideally have the same dimension as the parameter to be estimated. This result is supported by theoretical findings in \cite{li2018asymptotic}.
\cite{li2018asymptotic} and \cite{frazier2018asymptotic} have considered the asymptotic properties of ABC, with ABC tolerances decreasing as the amount of information in the data goes to infinity. Both papers have shown that convergence of ABC posteriors depends on the relationship between the rate of convergence of the summary statistics and that of the tolerance and they  have proved results on the asymptotic shape of the ABC posterior distribution. These results again depend on the relationship between the rate of convergence of the summary statistics and that of the tolerance. In particular, posterior consistency is only proved in the case where all summary statistics converge at a rate that is much faster than that of the tolerance. The shape of the asymptotic ABC posterior distribution is only proved in situations where all dimensions of the summary statistics converge at the same rate.

In this work, we extend the results of \cite{frazier2018asymptotic} to the case where different components of the summary statistics converge at different rates, with some possibly not converging at all. We first prove consistency of the ABC posterior where different components of the summary statistics are allowed to converge at heterogeneous rates. We next prove a general result on the asymptotic shape of the ABC posterior in the same context and our results cover the more realistic case where certain summary statistics do not converge at all. 

A well known technique to reduce the curse of the dimension of the set of summary satistics is based on non linear regressions, typically a post processing step, as introduced by \cite{Blum_2009}, see also \cite{blum2010approximate}. Recently \cite{li2018convergence} have shown, in the special case of asymptotically normal summary statistics concentrating at the same rate, that the local linear postprocessing step proposed in \cite{Blum_2009} leads to a significant improvement in the theoretical behaviour of the ABC posterior. 
However, the post-processing step is in general aimed at reducing the impact of the dimension of the summary statistics: it is therefore important to study its efficiency in a context where the summary statistics are not as well behaved as considered in \cite{li2018convergence}. In this paper we fill this gap by showing that local linear post-processing induces significant improvement even when summary statistics have heterogeneous behaviour. 

 In Section \ref{section assumptions} we provide details of our set-up, state the assumptions that we will be using and introduce key notation. In Section \ref{section theorems} we state our result on the asymptotic form of the ABC posterior and in Section \ref{sec:loclinear} we study its consequence on the local linear regression post-processing strategy. In Section \ref{section simulation study} we illustrate these results empirically. A short discussion is provided in Section \ref{section discussion}. The details of proofs of theoretical results are left to the appendix.

\section{Background} \label{section assumptions}

We observe data $y \in \mathbb{R}^n$ and assume that they arise from the model $\{P_{\theta}(\cdot):\theta \in \mathbb{R}^d \},$ where $P_{\theta}(\cdot)$ is a  density function.  We denote by $\theta_0 \in \mathbb{R}^d$ the unknown true value of interest that generated the observed data $y.$ We denote by $\pi(\cdot)$ the prior density on parameter space, and by $\Pi(\cdot)$ the corresponding cumulative density function.

The idea of ABC is to make inference on the posterior distribution using Monte Carlo samples of parameter-data pairs $(\theta_i,z_i) \in \mathbb{R}^d \times \mathbb{R}^n,$ simulated from the forward model. Distances between observed data $y$ and simulated data $z_i$ determine the role $\theta_i$ will play in the estimation of $\theta_0.$ 

When the dimension of data $n$ is large, it is inefficient to compute distances on the raw data.
 It is thus common practice to instead compute distances between lower dimension summary statistics of the observed and simulated data. We thus define a summary function $\eta: \mathbb{R}^n \rightarrow \mathbb{R}^k$ from data space to summary space, where $k < n$. Summary statistics will typically be sample moments or quantiles of the data, although many other summary statistics have been considered in the literature. In this work, we consider the Euclidean distance.

Although more sophisticated ABC algorithms now exist, we will focus on the simple accept/reject ABC algorithm \citep{tavare1997inferring, pritchard1999population}, described in Algorithm \ref{algo1} below. Algorithm \ref{algo1} generates a \textit{reference table} from the model consisting of (parameter, summary statistic) pairs. For a given tolerance level $\epsilon,$ parameters corresponding to data within a distance $\epsilon$ of the observed data are accepted. Parameters corresponding to data farther away than $\epsilon$ from the observed data are rejected. The accepted values form a sample of the ABC posterior distribution which is then used to estimate quantities of interest by Monte Carlo.

\begin{algorithm}
\caption{ Accept/reject ABC} \label{algo1}
\begin{algorithmic}
\Require Observed data $y$, summary statistics $\eta$, threshold $\epsilon$
\For{$i = 1,\ldots,M$}
    \State Simulate $\theta^i \sim \Pi(\cdot)$
    \State Simulate $z^i = \left( z_1^i,\ldots, z_n^i \right) \sim P_{\theta^i}(\cdot)$
    \If{$\norm{\eta(y)-\eta(z^i)}\leq \epsilon$}
    	\State Accept $\theta_i$
	\EndIf
\EndFor
\end{algorithmic}

\end{algorithm}

\cite{frazier2018asymptotic} and \cite{li2018asymptotic} suggest that the tolerance $\epsilon$ should depend on $n$, the dimension of the data, and should tend to zero as $n$ goes to infinity. Indeed, as $n$ increases, information about underlying parameters accumulates in samples. If a simulated parameter is close in distance to $\theta_0,$ then data generated from it should be close in distance to the observed data, $y.$ Hereafter, we will thus denote the ABC tolerance by $\epsilon_n,$ and let $\lim_{n \rightarrow \infty}\epsilon_n = 0.$

Defining an approximation to the likelihood as
\begin{align}
\tilde{p}_{\epsilon_n,\theta}\left( \eta(y) \right) := \int  \bm{1}_{ \{\norm{\eta(y)-\eta(z)}\leq \epsilon_n \} } d P_{\theta} (z),
\label{e7}
\end{align}
this ABC accept/reject algorithm produces samples from the following pseudo-posterior distribution (when marginalizing out the simulated data $\eta(z)$):
\begin{align}
\pi_{\epsilon_n}(\theta) \propto \pi(\theta) \tilde{p}_{\epsilon_n,\theta}\left( \eta(y) \right).
\label{e8}
\end{align}

 We will show that properties of the asymptotic ABC posterior depend on the relationship between $\epsilon_n$ and the rate at which summary statistics converge to some well-defined limit. We formalize the notion of convergence of summary statistics in Assumption \ref{assump1} below.

\begin{assumption}
There exists some Lipschitz continuous mapping $b: \mathbb{R}^d \rightarrow \mathbb{R}^k$ such that, for all $1 \leq j \leq k,$ there exists some sequence $v_{nj}$ such that
\begin{align*}
&v_{nj}\left( b(\theta)_j - \eta(z)_j \right) = O_P(1).
\end{align*}
Without loss of generality, we assume $v_{n1}\leq v_{n2} \leq \ldots \leq v_{nk}.$
\label{assump1}
\end{assumption}

In this work, we consider the novel setting where, for some $1 \leq k_0 < k,$ the statistics from 1 to $k_0$ converge at slow rates and the statistics from $k_0+1$ to $k$ converge at fast rates, i.e.
\begin{align*}
\lim_{n \rightarrow \infty} v_{nj}\epsilon_n = 0  \qquad & \forall 1 \leq j \leq k_0\\
\lim_{n \rightarrow \infty} v_{nj}\epsilon_n = \infty \qquad & \forall k_0 +1\leq  j \leq k.
\end{align*}

This generalizes the setting considered by \cite{li2018asymptotic} and \cite{frazier2018asymptotic}, who proved posterior consistency in the special case of a single rate $v_{n1}=\ldots=v_{nk}$.

Throughout this chapter, for any vector $g,$ we let $g_{(1)}=(g_1,\ldots,g_{k_0})$ denote the vector of length $k_0$ of all components $g_j$ of $g$ with $j \leq k_0$, and we will refer to these components as the slow components. We let $g_{(2)}=(g_{(k_0+1)},\ldots,g_k)$ denote the vector of length $(k-k_0)$ of all components $g_j$ of $g$ with $j \geq (k_0+1)$, and we will refer to these components as the fast components. We also write $D_n = \text{diag}(v_{n,j}, j\leq k)$; $D_{n,1}$ its upper $k_0$ sub-matrix and $D_{n,2}$ its lower $k-k_0$ sub-matrix. Set $Z_{n}(\theta)$ the random vector  $ D_n (\eta(z) - b(\theta))$ with $z \sim P_\theta$, so that $Z_{n,j}(\theta) = v_{n,j}(\eta_j(z)-b_j(\theta))$. 
%We also write $\Sigma_{n,1} = \text{diag}(v_{n1}, \cdots, v_{nk_0})$, the diagonal matrix with elements $v_{nj}$ on the diagonal. We write similarly $\Sigma_{n,2} = \text{diag}(v_{n(k_0+1)}, \cdots, v_{nk})$

In addition to Assumption \ref{assump1}, our results rely on the following assumptions, which we discuss in Remarks \ref{remark1} and \ref{remark2}. We show that these assumptions are verified on a toy example in Example \ref{eg1}.

\begin{assumption}
The matrix $\nabla b_{(2)}(\theta_0)$ is of full rank, where $\nabla$ represents the gradient operator.
\label{assump2}
\end{assumption}

\begin{assumption}
There exist some constant $\delta>0,$ some strictly positive bounded Lipschitz continuous function $\gamma: \mathbb{R}^{k_0}\rightarrow \mathbb{R}^+,$ some constant $R>0,$ and some $o(1)$ sequence $L_n$ such that, for all compact sets $K \in \mathbb{R}^{k_0},$ and for all $1\geq t>0$,
\begin{align*}
\sup_{\norm{\theta-\theta_0}<\delta} \sup_{m \in K} \left| \frac{P_{\theta} \left( \sum_{j=1}^{k_0} \frac{(Z_{n,j}(\theta) - m_j)^2}{v_{n,j}^2}   <  t^2\epsilon_n^2 \right) }{t^{R} L_n} - \gamma(m) \right| = o(1).
\end{align*}
\label{assump3}

\end{assumption}

\begin{assumption}
There exist $\kappa>d$ such that $v_{n,k_0+1}^{-\kappa}=o(L_n\epsilon_n^d)$, and there exists some function $c: \Theta \rightarrow \mathbb{R}^+$ verifying $\int_{\Theta} c(\theta) \pi(\theta)d\theta < \infty$, such that the following holds for all $n\geq1$, for all $u>u_0$ and for all $\theta \in \Theta$:
\begin{align*}
\max_{j\geq k_0+1}P_{\theta} \left( |Z_{n,j}(\theta)|  > u \right) \leq  c(\theta) u^{-\kappa}.% ; \quad\int_{\Theta} c(\theta) \pi(\theta)d\theta < \infty.
\end{align*}

\label{assump4}
\end{assumption}

\begin{assumption}
For all $M>0$, there exist  %monotone nonincreasing functions 
$M_2>0$, $\kappa >d$ and a non decreasing  sequence $u_n\geq u_0>0$  verifying $u_n = o( v_{n,k_0+1} \epsilon_n) $, such that  for all $n $ large enough,   $u\geq u_n$, $\|\theta - \theta_0\| \leq \delta$, and  $j > k_0$,
\begin{align*}
 &  \sup_{\|m\|\leq M} P_{\theta}\left(  |Z_{n,j}(\theta)|  > u | Z_{n,(1)}(\theta) = m \right)  \leq M_2 u^{-\kappa}  , 
\end{align*}
with 
$$\left(\frac{  v_{n,k_0} }{ v_{n,k_0+1} }\right)^{\kappa -d} (\epsilon_n v_{n,k_0+1})^{-d}= o(L_n). $$
% \ref{assump3}.
%Furthermore, the function $\bar{\rho}(\cdot)$ is such that there exists some (slowly) monotone increasing sequence $M_n$ which satisfies $M_n=o(\sqrt{\epsilon_n v_{n(k_0+1)}})$ and which satisfies $\bar{\rho}(M_n)=o(\epsilon_n^d).$
\label{assump5}
\end{assumption}

\begin{remark}
Assumption \ref{assump2} guarantees that enough summary statistics have a fast rate of convergence. Assumption \ref{assump3} guarantees sufficient stability of the summary statistics with slow rates of convergence. Typically, $L_n$ will be a sequence proportional to $\prod_{i=1}^{k_0} v_{ni} \epsilon_n$ and $R$ will be equal to $k_0.$ (See Example \ref{eg1} below). Assumption \ref{assump4} controls the tail behavior of the summary statistics with fast rates of convergence.  Assumption \ref{assump4} is similar to  the tail assumption of \cite{frazier2018asymptotic} and we refer to  their discussion on this assumption. It holds true in particular if $\limsup_n E_\theta(\|Z_{n,(1)}\|^\kappa)<\infty$. 
 Assumption \ref{assump5} is implied by Assumption \ref{assump3} and Assumption \ref{assump4} in the particular case where the vectors of slow and fast converging summary statistics are mutually independent, but holds more generally than that.
\label{remark1}
\end{remark}

\begin{remark}
Our Assumptions are slightly weaker than those of \cite{li2018asymptotic} and of \cite{frazier2018asymptotic}. While in Assumption \ref{assump1} we require just stochastic boundedness, these authors require central limit theorems for the summary statistics. More importantly we do not impose, as in \cite{li2018asymptotic} that the summary statistics concentrate at the same rate nor do we impose some limiting distribution for the fast ones. However $\gamma$ in Assumption \ref{assump3} can be thought of as the limiting density of the the slow (first $k_0$) summary statistics. Compared to \cite{frazier2018asymptotic}, we do not impose that $\epsilon_n$ is either smaller or larger than all summary statistics. 
 While our Assumption \ref{assump3} and Assumption \ref{assump4} apply only to the summary statistics with slow and fast rates of convergence respectively, \cite{frazier2018asymptotic} make similar assumptions for the full vector of summary statistics.
\label{remark2}
\end{remark}

\begin{example}\label{eg1}
We verify all of the assumptions for a simple uniform toy example. We assume that the data are iid from a continuous unit uniform distribution with unknown location parameter $\theta \in \mathbb{R}:$ $z_i\sim \mathcal U\left(\theta-\frac{1}{2},\theta+\frac{1}{2} \right), i \in \{1,\ldots,n\}.$ We put a standard uniform prior on the parameter: $\theta \sim \mathcal U(0,1).$ We take the first $k_1$ summary statistics to be the first $k_1$ observations. These do not converge, so we have $b(\theta)_i=0$ and $v_{ni}=1$ for $i \in \{1,\ldots,k_1 \}$. We take $\eta_{k_1+1}(z) = \bar z_n = \frac 1n\sum_{i=1}^nz_i$ and $\eta_{k_1+2}(z) =  \max_{ i \leq n}z_i $. The statistic $\eta_{k_1+1}(z)$ converges at the rate $v_{n,k_1+1}=\sqrt{n}$ to $b_{k_1+1}(\theta)=\theta$ and the statistic $\eta_{k_1+2}(z)$ converges at the rate $v_{n,k_1+2}=n$ to $b_{k_1+2}(\theta)=1/2+\theta$.  We take $\epsilon_n = o(1)$ and we consider two scenarii.  Scenario 1: $\frac1n <<\epsilon_n<<\frac{1}{\sqrt{n}}$; Scenario 2:  $\frac{1}{\sqrt{n}} <<\epsilon_n<< 1$.

 In Scenario 1 $k_0=k_1+1$ and $\eta_{(1)}(z) = \left( z_1,\ldots,z_{k_1}, \bar z_n\right) \in \mathbb{R}^{k_1+1 }$ while in scenario 2,  $k_0=k_1$ and $\eta_{(1)}(z) = \left( z_1,\ldots,z_{k_1}\right) \in \mathbb{R}^{k_1}$.  Assumptions \ref{assump1} and \ref{assump2} are trivially verified in both scenarii. 
 Using the same notation as before, we are then in the setting where $\forall i\leq k_0,\lim_{n \rightarrow \infty} v_{ni}\epsilon_n=0$  and $\lim_{n \rightarrow \infty} v_{n(k_0+1)}\epsilon_n = \infty.$  
 
We now prove that assumptions \ref{assump3}-\ref{assump5} are verified. We treat the case of scenario 1, which is more difficult.

Let $K_1$ be a compact subset of $(-\frac12, \frac12)^{k_1} $ and $K=K_1 \times [-M,M]$. We have $Z_{n,(1) }= (z_1-\theta, \cdots, z_{k_1} - \theta, \sqrt{n}( \bar z_n - \theta) )$ and $\bar Z_{n,(2)} = n\left(\eta_{k_1+2} - \theta-\frac12\right)$. Note that $\sqrt{n}( \bar z_n - \theta) $ converges to a $\mathcal N(0,\frac{1}{12})$ distribution; let $\varphi$ be the density function of that limit distribution. Let $m\in K$. Then:
\begin{align*}
&P_{\theta} \left( \norm{ (Z_{n,(1)}(\theta)-m } \leq t \epsilon_n \right) = E_\theta\left( P_{\theta}\left( (\bar z_n-m_{k_1+1})^2 \leq t^2 \epsilon_n^2  - \sum_{j=1}^{k_1}(Z_{n,j}-m_j)^2 | z_1, \cdots, z_{k_1}\right) \right)\\
&= E_\theta \left( \1_{\sum_{j=1}^{k_1}(Z_{n,j}-m_j)^2\leq t^2 \epsilon_n^2 } \cdot 2 \sqrt n \sqrt{t^2 \epsilon_n^2  - \sum_{j=1}^{k_1}(Z_{n,j}-m_j)^2}\varphi(m_{k_1+1})(1 + o(1) )\right)\\
& = 2\varphi(m_{k_1+1})\sqrt n\prod_{j=1}^{k_1}(f_{\theta}(m_j)+o(1))  \int \1_{\sum_{j=1}^{k_1}(z_j-m_j)^2\leq t^2 \epsilon_n^2 }\sqrt{t^2 \epsilon_n^2  - \sum_{j=1}^{k_1}(z_j-m_j)^2}dz_1\cdots dz_{k_1} \\
&= 2 \varphi(m_{k_1+1})\sqrt n\prod_{j=1}^{k_1}(f_{\theta}(m_j)+o(1))(t\epsilon_n)^{k_1+1}\int_0^1 u^{k_1}\sqrt{1-u^2}du
\end{align*}
and Assumption \ref{assump3} is verified, with $R=k_1+1=k_0$, $L_n = 2\sqrt n\epsilon_n^{k_1+1}\int_0^1 u^{k_1}\sqrt{1-u^2}du,$ and $\gamma(m) = \varphi(m_{k_1+1})\prod_{i=1}^{k_1} f_{\theta}(m_i)$. 

We also have 
\begin{align*}
&P_{\theta}\left( \left|n\left(\eta_{k_1+2}-\theta-\frac12\right) \right|  \geq u \right)=Pr \left( \max_i z_i\leq \theta + \frac12 -  \frac u n \right)= \left(1-\frac u n\right)^n \leq e^{-u}
\end{align*}
so that Assumption \ref{assump4} holds for all $\kappa >0$. 

Here $Z_{n,(2)}$ and $Z_{n,(1)}$ are not independent, but we can verify Assumption \ref{assump5}. 
First note that, with $z_{(n)} = \max_i z_i$ and for all $m \in \mathbb R^{k_1+1}$, writing 
\begin{align*}
P_\theta \left(  n| z_{(n)}-\theta-\frac12 | > u  \mid Z_{n,(1)}=m\right) &= P_\theta \left( z_{(n)}\leq \theta +\frac12 -\frac un \mid Z_{n,(1)}=m\right) \\
&\leq  P_\theta \left( \max_{i\geq k_1+1} z_i\leq \theta +1/2 -u/n \mid Z_{n,(1)}=m\right) \\
&= P_\theta \left( \max_{i\geq k_1+1} z_i\leq \theta +\frac12 -\frac un \mid \bar z_n -\theta = \frac{m_{k_1+1}}{\sqrt{n}}  \right)  
\end{align*}
Hence without loss of generality we can work with $k_1=0$. Consider the change of variables $ x_i = z_i-\theta$ for $i\geq  2$ and $x_1 = \bar z_n-\theta$, whose distribution is independent of $\theta$. The joint density of 
$x= (x_1, \cdots, x_n)$ is given by 
$$ f_x( x ) = \1_{nx_1- (n-1)\bar x_{n-1} \in (-\frac12, \frac12)}\prod_{i\geq 2} \1_{x_i\in(-\frac12,\frac12)}  \qquad \text{with }\bar x_{n-1} = \frac{1}{n-1}\sum_{i\geq 2} x_i.$$
This leads to for all $m\in \mathbb R$, 
\begin{align}\label{maxmean}
P \left( \max_{i\geq 2 }x_i \leq \frac12 -\frac un \mid x_1 = \frac{m}{\sqrt{n}}\right) & = \frac{ \int_{[-\frac12,\frac12-\frac un]^{n-1}}\1_{\sqrt{n} m - (n-1)\bar x_{n-1} \in (-\frac12, \frac12)}dx_2\cdots dx_n }{ \int_{[-\frac12,\frac12]^{n-1}}\1_{\sqrt{n} m - (n-1)\bar x_{n-1} \in (-\frac12, \frac12)}dx_2\cdots dx_n}
\end{align}

We also have that 
\begin{align*}
\int_{[-\frac12,\frac12]^{n-1}}\1_{\sqrt{n} m - (n-1)\bar x_{n-1} \in (-\frac12, \frac12)}dx_2\cdots dx_n = P\left( \left|\sqrt{n-1} \bar x_{n-1}-m\sqrt{\frac{n}{n-1}} \right| \leq \frac{1}{2\sqrt{n-1}} \right)\asymp \frac{ e^{- 6 m^2 } }{ \sqrt{n} },
\end{align*}
which plugged into \eqref{maxmean}  implies 
\begin{align*}
P \left( \max_{i\geq 2 }x_i \leq \frac12 -\frac un \mid x_1 = \frac{m}{\sqrt{n}}\right) & \lesssim  \sqrt{n}e^{6 m^2 }  \int_{[-\frac12,\frac12-u/n]^{n-1}}dx_2\cdots dx_n \\
&\lesssim \sqrt{n}e^{6 m^2} e^{-u} \lesssim e^{6 m^2} e^{-u/2} \quad \forall u \geq \log n:=u_n
\end{align*}
Therefore as soon as  $ \epsilon_n >>\frac{\log n}{n}$, Assumption \ref{assump5} holds true for all $\kappa >0$.

\end{example}

\section{Main theorems} \label{section theorems}

Our first result, Theorem \ref{theorem1} below is a Bayesian consistency result. It asserts that the ABC posterior density of any set which does not include the parameter which generated the observations, $\theta_0$, behaves like an $o_P(1)$ random variable.  Since the ABC posterior will differ from the true posterior given the observations, such a result is crucial if one wishes to quantify uncertainty based on the ABC posterior.  

\begin{theorem}
Under Assumptions \ref{assump1}, \ref{assump2}, \ref{assump3}, and \ref{assump4} our ABC posterior distribution concentrates: There exists some monotone decreasing sequence $\lambda_n$ with $\lambda_n \rightarrow 0$ such that
\begin{align*}
\int_{\norm{\theta-\theta_0} \geq \lambda_n} \pi_{\epsilon_n}(\theta| \eta(y)) d \theta = o_P(1).
\end{align*}
\label{theorem1}
\end{theorem}

The rate of concentration of the ABC posterior, $\lambda_n,$ is of the same order as the sequence $\bar{\lambda}_n$ of Assumption \ref{assump4}. Thus, following Remark \ref{remark1} on the form of the function $\rho(\cdot,\cdot)$ and the sequence $L_n$, we typically will have that, the faster the rate of the convergence of the fast statistics, $v_{n (k_0+1)}$, the faster the rate $\lambda_n$ will be. The greater the quantity of slow converging statistics, $k_0$, and the slower the slow converging statistics converge, the slower the rate $\lambda_n$ will be.

Our second result, Theorem \ref{theorem2} completely characterises the shape of the ABC posterior.

\begin{theorem}
Under Assumptions \ref{assump1}, \ref{assump2}, \ref{assump3}, \ref{assump4} and \ref{assump5}, the asymptotic ABC posterior may be expressed in closed form as 
\begin{align}
\pi_{\epsilon_n}(\theta | \eta (y)) \propto \mathbbm{1}_{\{ \norm{\nabla b_{(2)}(\theta_0) (\theta - \theta_0)} \leq \epsilon_n \} } \left( 1 - \frac{ \norm{\nabla b_{(2)}(\theta_0) (\theta - \theta_0)}^2}{\epsilon_n ^2} \right)^{\frac{R}{2}},
\label{th2}
\end{align}
where $R$ is the positive constant defined in Assumption \ref{assump3}.
\label{theorem2}
\end{theorem}

As discussed in Remark \ref{remark1}, $R$ will typically be equal to $k_0,$ the number of summary statistics which converge at the slow rate. In the special case where $R=0,$ the shape of the ABC posterior distribution simplifies to a uniform distribution over the ellipsoid $\{\theta: \norm{\nabla b_{(2)}(\theta_0)(\theta-\theta_0)}\leq \epsilon_n \}.$ This is consistent with results in \cite{frazier2018asymptotic} and in \cite{li2018convergence} for the case where all statistics converge at the fast rate (i.e. where $k_0=0$).

Redundant summary statistics which do not converge at all play exactly the same role on the shape of the asymptotic ABC posterior as summary statistics which converge at the slow rate.

The larger $R$ is, the more concentrated the theoretical mass of \eqref{th2} will be around $\theta_0$. However, we will see in Lemma \ref{lemma3} that large $R$ leads to low acceptance rate in Algorithm \ref{algo1}, and thus high Monte Carlo error. 

Interestingly, the number of summary statistics which converge at the fast rate (i.e. $k-k_0$) will have no impact on the the rate of posterior concentration nor on the the shape of the ABC posterior (beyond the requirement $(k-k_0)>d$ by Assumption \ref{assump2}).

\section{Local linear regression correction} \label{sec:loclinear}

ABC practitioners routinely use post-processing to improve the quality of the pseudo-posterior. Beaumont2002 introduced the idea of a local linear regression on the ABC output; empirical studies have since shown that this post-processing step can vastly ameliorate the pseudo-posterior, for a negligible computational overhead. In this section, we give results on the asymptotic behaviour of the regression-adjusted pseudo-posterior. 

In general, post-processing corrections use the following idea: Consider the  \textit{pseudo model}
$$\theta = m(S) + u, \quad (\theta, S) \sim \pi_{\epsilon_n}(\theta, dS) \propto \pi(d\theta) P_\theta(dS)\1_{|S-S_0\leq \epsilon_n} $$ 
and samples $( \theta_t,  S_t) $ from the above distribution. Our distribution of interest is the distribution of $m(S_0)+u$. To approximate it, we learn the model $m$ and consider the residuals 
$\hat u_t = \theta_t - m(S_t)$;  the corrected ABC posterior samples are:
$$ \theta_t = \theta_t + m(S_0) - m(S_t) .$$ 

The regression adjustment we consider here corresponds to the locally linear model case,   where $m(S) = (B,\,  \beta_0)^T (S, \, 1)$ so that the targeted $B \in \mathbb R^{d\times k}, \beta_0\in mathbb R^d$ minimizes 
\begin{equation}\label{contrast}
L(B, \beta_0) = E_n[\|\theta-\theta_0 - \beta_0 - B^T ( S-S_0)\|^2] 
\end{equation}
Let $\beta(j)$ the $j$th row of $B$, $\beta_{(1)}(j)$ (resp. $\beta_{(2)}(j)$) the first $k_0$ components of $\beta(j)$ (resp. the last $k-k_0$) . 

In addition to the assumptions \ref{assump1}-\ref{assump5} we also assume the following: 
\begin{itemize}
\item [A1] The function $\theta \rightarrow E_\theta ( Z_n(\theta)) $ is locally Lipschitz on a neighbourhood of $\theta_0$ with Lipschitz constant $L$. 
%\item [A1] There exists $\epsilon>0$ such that $\theta \rightarrow E_\theta ( Z_n(\theta)) $ is Lipschitz near $\theta_0$ with Lipschitz constant $L$. 

\item [A2] There exists $\epsilon>0$ such that 
$$ \sup_{\|\theta - \theta_0\|\leq \epsilon} E_\theta\left ( \|Z_{n}(\theta)\|^4 \right) < \infty. $$ 
\end{itemize}

We then have the following theorem.
\begin{theorem}\label{th:LS}
Under assumptions \ref{assump1}-\ref{assump5} and assuming that [A1] and [A2] above hold, then on a set of $y$ whose probability goes to 1, any minimizer in $B$ of $L(B, \beta_0)$ verifies 
\begin{equation}\label{minB}
 C_{11} D_{n,1}^{-1} \beta_{(1)}(j) = O(\epsilon_n/v_{n,1}), \quad \nabla b_2(\theta_0)^T \beta_{(2)}(j) = e_j+ O(1/v_{n,1}),
 \end{equation}
 where $e_j $ is the $j$-th vector in the canonical basis of $\mathbb R^d$ and 
 $$ C_{1,1} = E_n ( (Z_{n,(1)}(\theta) - E_n(Z_{n,(1)}(\theta))) (Z_{n,(1)}(\theta) - E_n(Z_{n,(1)}(\theta)))^T) .$$ 

Let $B^*$ be the limit of $B(j)$ with minimal $L_2$ norm (by rows), then 
$B^*(j) = (0, \cdots , 0, \Gamma_2 e_j)$ with $\Gamma_2 = \nabla b_2(\theta_0)[\nabla b_2(\theta_0)^T\nabla b_2(\theta_0]^{-1} $.

Moreover 
if $$ \theta' = \theta - B^* (S-S_0 ),\quad \text{with} (\theta, S) \sim \pi_{\epsilon_n}(\theta, dS)$$
then 
$$ \theta_j' - \theta_{0j} =  e_j^T \Gamma_2D_{n,2}^{-1} (Z_{n,(2)}(\theta) -Z_{n,(2)}(\theta_0) ) + O_p(\epsilon_n^2 ) .$$ 
\end{theorem} 

An important consequence of Theorem \ref{th:LS} is that the oracle post-processing, i.e. the post processing associated to $B^*$, leads to a posterior contraction rate of order $\max( v_{n,k_0+1}, \epsilon_n^2)$. In comparison, Vanilla ABC leads to a posterior contraction rate of order $\max( v_{n,k_0+1}, \epsilon_n)$. The linear post-processing thus  corresponds to what would be obtained if $\epsilon_n$ was replaced by $\epsilon_n^2$, without increasing the order of the computational cost . This shows the importance of the post-processing as a tool towards dimension reduction, and interestingly the local linear approach already leads to a significant theoretical  improvement, even in the general  and more realistic framework of summary statistics which have different concentration properties. 
The proof of Theorem \ref{th:LS} is provided in Section \ref{pr:th:LS}.

\section{Simulation study} \label{section simulation study}
\subsection{Vanilla ABC}

We perform accept/reject ABC to obtain Monte Carlo samples from the ABC posterior of a variation of Example \ref{eg1}. All of the assumptions of Section \ref{section assumptions} are satisfied, and so, by Theorem \ref{theorem2}, the asymptotic shape of the ABC posterior is available in closed form. The goal of this simulation study is to provide empirical support to this theoretical result.

We recall the data distribution, the prior distribution, and the summary statistics used in Example \ref{eg1}. Data are distributed according to a continuous unit uniform distribution with unknown location parameter $\theta_0 \in \mathbb{R}: y_i \sim U(\theta-\frac{1}{2},\theta+\frac{1}{2}) \hspace{0.3cm} \forall i \in \{1,2,\ldots,n \}.$ We put a uniform prior on the parameter: $\theta \sim U(0,1).$ We use the following vector of summary statistics.
\begin{align*}
\eta(y)= \left(y_1,\ldots,y_{k_0},  \left(\max_{ (k_0+1)\leq i \leq n} y_i + \min_{(k_0+1) \leq i \leq n} y_i \right)/2  \right)
\end{align*}

We have $b(\theta)_i=\theta$ and $v_{ni}=1$ for all $1 \leq i \leq k_0$ and we have $b(\theta)_{k_0+1}=\theta$ and $v_{n(k_0+1)}=n$. We set the tolerance to be $\epsilon_n=\frac{C}{\sqrt{n}}$ where $C$ is some constant. We are then in the regime where $\lim_{n \rightarrow \infty} v_{ni}\epsilon_n =0 \hspace{0.3cm} \forall i \leq k_0,$ and $\lim_{n \rightarrow \infty} v_{n(k_0+1)}\epsilon_n = \infty.$ We use the Euclidean norm for distances.

By Lemma \ref{lemma3} we have that this choice for the sequence $\epsilon_n$ is equivalent to setting the sequence of ABC acceptance probabilities to be as follows.
\begin{align}
\alpha_n \propto L_n \epsilon_n^d \propto \epsilon_n^{k_0+d} \propto n^{-\frac{k_0+1}{2}}
\label{alpha}
\end{align}

By Theorem \ref{theorem2}, we have the following closed-form expression for the ABC posterior for this example, where $C$ is the constant which satisfies $\epsilon_n = \frac{C}{\sqrt{n}}.$

\begin{align}
\pi_{\epsilon_n}(\theta | \eta(y)) \propto \mathbbm{1}_{\{ \norm{\theta-\theta_0}<\frac{C}{\sqrt{n}} \}} \left( 1 - \frac{n \norm{\theta - \theta_0}^2}{C^2} \right)^{\frac{k_0}{2 }}
\label{theoretical}
\end{align}

We run this experiment with $k_0 = 2$ and for $n\in\{10^4, 10^5, 10^6\}$. The results, shown in Figure \label{fig:conv2}, show excellent agreement between the empirical posterior and the form predicted by Theorem \ref{theorem2}.

\begin{figure}
\includegraphics[height=.25\textheight]{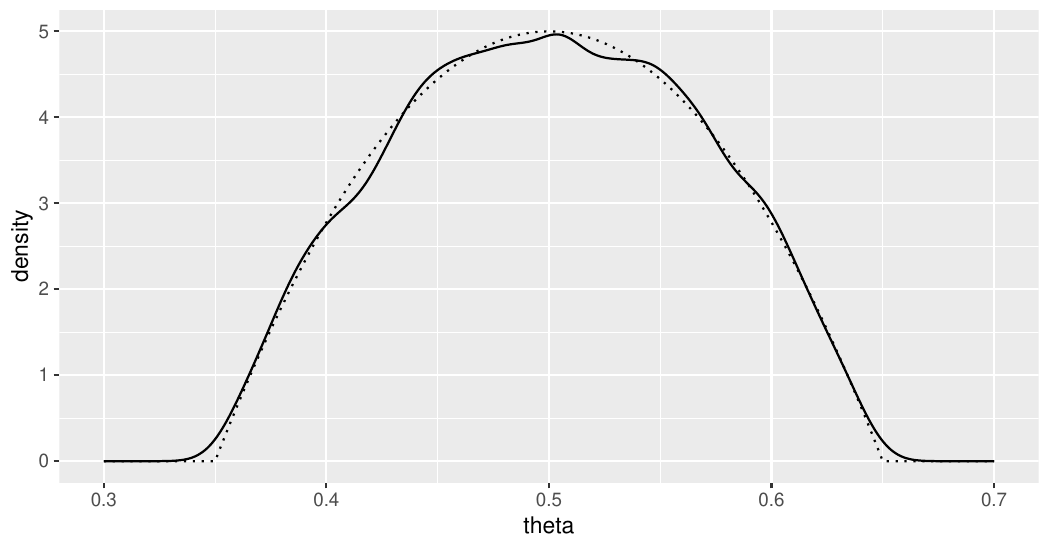}
\includegraphics[height=.25\textheight]{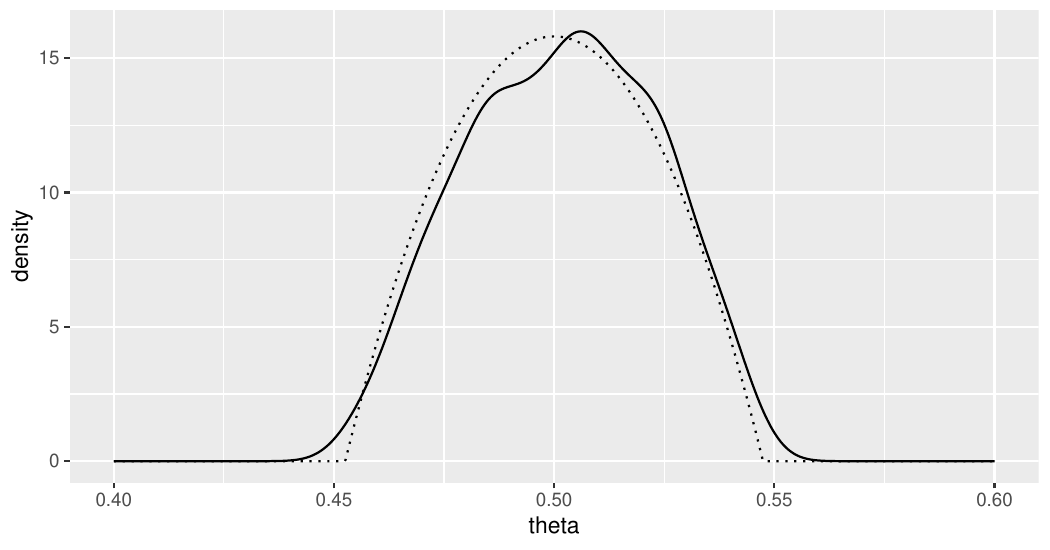}
\includegraphics[height=.25\textheight]{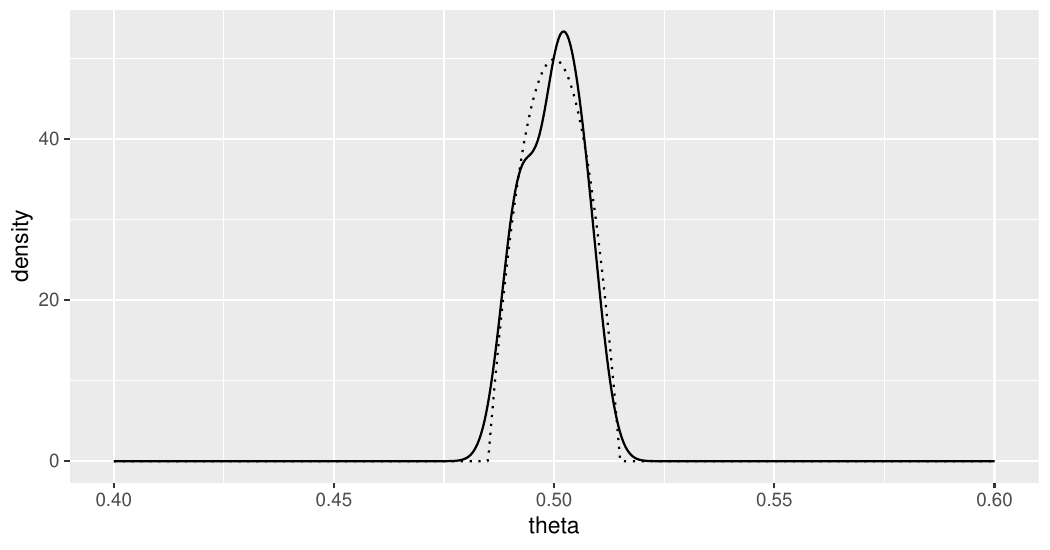}
\caption{ABC posterior for the uniform example, using 2 non converging statistics and 1 fast statistic. Top to bottom: $n=10^4, 10^5, 10^6$. Full line: empirical posterior; dashed line: theoretical posterior as predicted by Theorem \ref{theorem2}.}
\label{fig:conv2}
\end{figure}

%
%We perform 3 experiments, one for each of $k_0=1, 5, 10$, and let $n$ take values in the range $\{10^5,2(10^5),3(10^5),4(10^5),5(10^5)\}.$ In each case, we ensure that $\alpha_n=0.3$ for $n=10^5$. When $n$ increases, $\alpha_n$ decreases accordingly.
%
%Figure \ref{fig1} (left) shows the shapes of the empirical ABC posterior distributions for $k_0=1.$ Different coloured curves indicate different values of $n.$ Figure \ref{fig1} (right) shows the shapes of the corresponding theoretical ABC posterior distributions of \eqref{theoretical}. Figures \ref{fig5} and \ref{fig10} show the same as the above, with $k_0=5$ and $k_0=10$ respectively. The similarity between the shapes corresponding to the simulated and to the theoretical ABC posterior curves supports Theorem \ref{theorem2}.
%
%The reason why we chose $\alpha_n=0.3$ for $n=10^5$ regardless of $k_0$ was to keep the sizes of reference tables reasonable. If, alternatively, $\epsilon_n$ for $n=10^5$ were kept constant across different choices for $k_0,$ acceptance rates for large $k_0$ would be much smaller than acceptance rates for large $k_0.$ Low acceptance rates are costly in terms of computational time.
%
%It may seem counter intuitive that the ABC posterior for $k_0=1$ is more concentrated than the ABC posterior for $k_0=5$ and for $k_0=10$ for fixed $n.$ We emphasise that a fixed acceptance probability implies smaller $\epsilon_n$ for small $k_0$ and larger $\epsilon_n$ for large $k_0.$ ABC posteriors must therefore not be compared accross different values of $k_0$ in this study.

\subsection{Simulations with postprocessing}\label{sec:sim_post}

We return to the example of estimating the location parameter of uniform observations. The data are iid $X_i\sim\mathcal U(\theta,\theta+1)$ distribution; we observe $n=10^4$ realizations. This time, we consider the statistics $S_1 = \frac{1}{\sqrt n}\sum_{i=1}^{\sqrt n} X_i$ and $S_2 = \min_{1\leq i\leq n}X_i$. The convergence rates are thus $n^{-1/4}$ for $S_1$ and  $n^{-1}$ for $S_2$.

We compute the posterior risk $E[(\theta-\theta_0)^2]^{1/2}$, where $\theta$ is drawn from the pseudo-posterior for decreasing values of $\epsilon_n$ both without and with post-processing. Recall that we expect the risk to decrease at rate $\epsilon_n$ in the Vanilla ABC case, but at rate $\epsilon_n^2$ with the post-processing, until the risk reaches a plateau when $\epsilon_n$ becomes smaller than $v_{n,k_0+1}=n^{-1}$. 

\begin{figure}
\includegraphics[width=.8\textwidth]{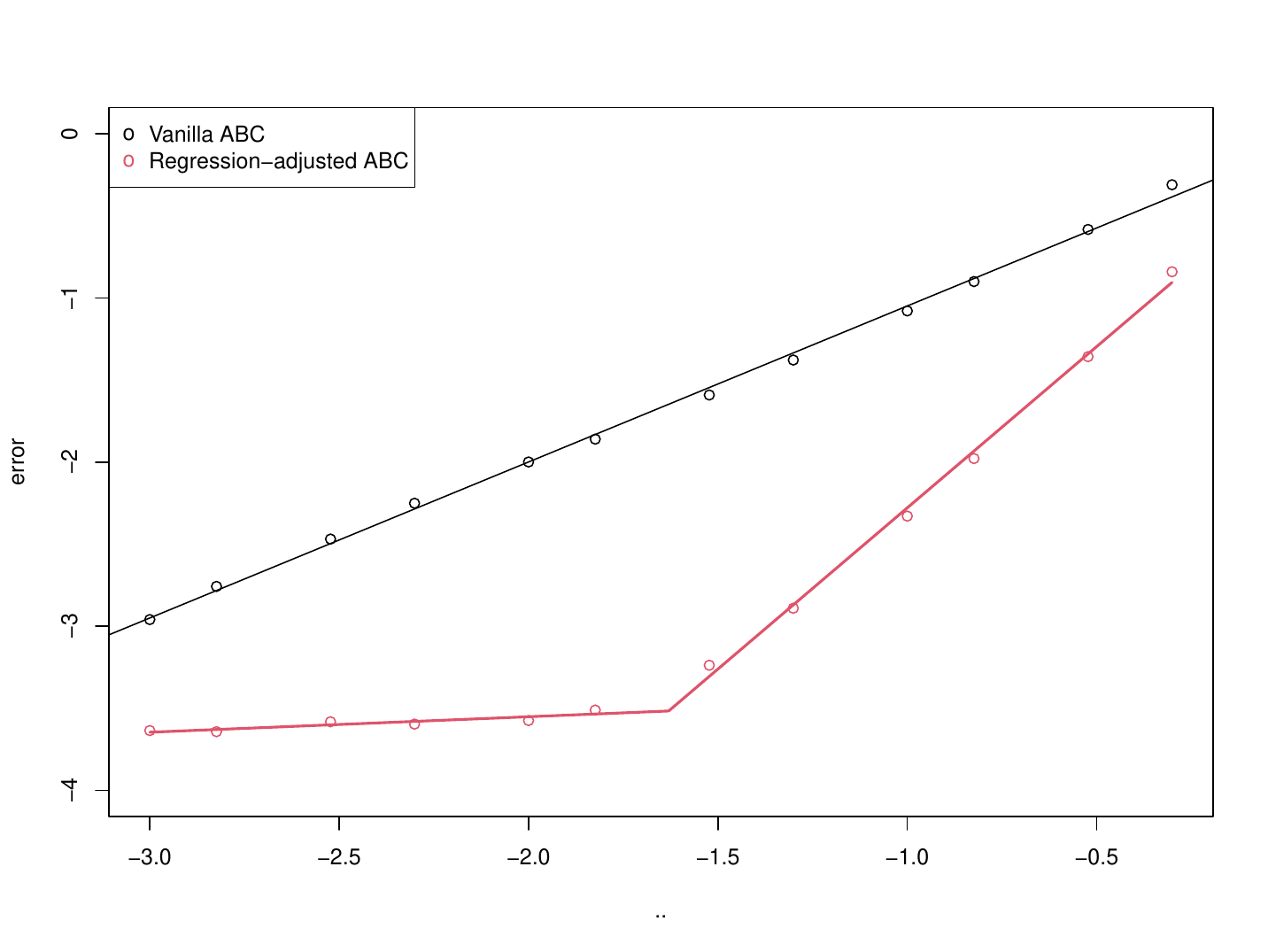}
\caption{Posterior risk for various values of $\epsilon$ for the example of Section \ref{sec:sim_post} (with post-processing correction), and fitted regression lines. Both $\epsilon$ and the posterior risk are shown on the log scale.}
\label{fig:sim_post}
\end{figure}

Figure \ref{fig:sim_post} shows the posterior risk on the log-log scale (the log is in base 10). Note that as expected, the posterior risk decreases when $\epsilon$ decreases. For the Vanilla ABC, the plateau is never reached for computational reasons. A linear regression estimates that in this example, the risk decreases at rate $\epsilon^\gamma$ with $\gamma=0.95$, very close to the theoretical value of $\gamma=1$. For ABC with post-processing, segmented regression (as implemented in the R package \texttt{segmented} \cite{muggeo2003estimating}) estimates that the risk decreases at rate $\epsilon^\gamma$ with $\gamma=1.87$, again close to the theoretical value of $\gamma=2$. With post-processing, the plateau is reached for $\epsilon\approx 10^{-1.6}$: there is thus no point in decreasing $\epsilon$ beyond this value, as we would lose Monte Carlo accuracy but not improve the accuracy of the pseudo-posterior.

\section{Discussion} \label{section discussion}

We prove posterior consistency, and provide a closed-form expression for the shape of the asymptotic ABC posterior distribution. Unlike in previous work, our results apply to the general case where different components of the summary statistics converge at different rates. In particular, we cover the case where certain components of the summary statistics do not converge at all. This set-up corresponds well to practical situations in applied statistics where large numbers of statistics are used, with potentially varying convergence rates.

Our theoretical proofs provide, as a byproduct, insight into the effect summary statistic choice and parameter dimension have on the Monte Carlo error.  By Lemma \ref{lemma3}, acceptance probability is directly proportional to the sequence $L_n$. As mentioned in Remark \ref{remark1}, $L_n$ will typically take the form $L_n = \prod_{i=1}^{k_0} \epsilon_n v_{ni}.$ Thus, we will typically have that the greater the number of slow summary statistics, the faster the acceptance probability will shrink to zero, and so the greater the Monte Carlo error will be. Lemma \ref{lemma3} also illustrates the curse of dimensionality, with acceptance probability decreasing rapidly for large parameter dimension $d.$ This observation is consistent with previous work \citep{fearnhead2012constructing}) which suggests making different estimations of subvectors of the vector of parameters separately.

In order for our results to hold true, at least $d$ summary statistics must be used that converge at the fast rate (Assumption \ref{assump2}). However, adding additional fast converging statistics (i.e. $ k-k_0>d$) will neither change the shape of the asymptotic ABC posterior nor increase the Monte Carlo error.

Despite the existence of more sophisticated methods, our theoretical results are limited to the basic accept/reject ABC algorithm (Algorithm \ref{algo1}). \cite{beaumont2010approximate} propose extensions, involving using general kernels and using regression adjustment. Instead of accepting all simulated $\theta^i$ for which $\norm{\eta(y)-\eta(z^i)}\leq \epsilon$ in step (3) of Algorithm \ref{algo1}, they propose accepting $\theta^i$ with probability $K(\eta(y),\eta(z^i))$, where $K(\cdot,\cdot)$ is some kernel. This way, parameters that generate data which is close to the observed data may be given a greater weight than parameters that generate data which is far away from the observed data. In addition, the author fits a regression model for simulated data-parameter pairs to correct for the difference between the simulated data and observed data in parameter estimation. It has been shown that such extensions can both reduce bias \citep{blum2010approximate}), and improve computational efficiency \citep{beaumont2010approximate}, \cite{li2018convergence}) in statistical analysis. Our next task will be to provide similar results to those presented in this work to the ABC methods of \cite{beaumont2010approximate}.

Throughout our proofs, we make the strong assumption that $\nabla b_{(2)}(\theta_0)$ is of full rank, i.e. sufficiently many summary statistics converge at the fast rate relative to the tolerance (see Assumption \ref{assump2}).  Asymptotic results on the ABC posterior in a more general setting where this assumption is lifted will be left to future research.

\newpage

\clearpage
\bibliographystyle{chicago} % apalike
\bibliography{biblio}

\newpage

\section{Appendix} \label{chapter appendix}

\subsection{Statements of lemmas} 
We consider the following sets: Let $\|\theta -\theta_0\|\leq \lambda_n $ and define
$w_n(\theta) = \|\nabla b_2(\theta_0)\frac{\theta-\theta_0}{\epsilon_n}\|^2$ together with 
\begin{align*}
A_n(\theta) & := \left\{ z; \norm{\eta_{(2)}(z) - b_{(2)}(\theta)}  \leq \frac{\delta_n \epsilon_n}{4}\left(1 \vee  \sqrt{w_n(\theta)}\right)\right\} \\
E_n(\theta) & := \{ z;\norm{\eta(y)-\eta(z)}\leq \epsilon_n \}\\
\tilde{E}_n (\theta)&:= \{z; \norm{\eta_{(1)}(y)-\eta_{(1)}(z)} \leq \epsilon_n \}\\
E_n'(\theta) &:= \{ z;\norm{\eta_{(1)}(y)-\eta_{(1)}(z)} \leq \epsilon_n \left( 1 - w_n(\theta)-\delta_n \right)^{\frac{1}{2}}  \} \quad \text{ if } \quad w_n(\theta)<1-\delta_n\\
E_n'' (\theta)&:= \{ z;\norm{\eta_{(1)}(y)-\eta_{(1)}(z)} \leq \epsilon_n \left( 1 - w_n(\theta) +\delta_n \right)^{\frac{1}{2}}  \}\\
\end{align*}
Obviously $\tilde E_n (\theta) \subset E_n (\theta) $ and $E_n'  (\theta)\subset E_n'' (\theta)\subset \tilde E_n (\theta)$ as long as $\delta_n < w_n(\theta)<1-\delta_n$.

\begin{lemma}
We can choose  $\lambda_n , \delta_n=o(1)$ such that the following inequalities hold: for all $M>0$,  $y \in \Omega_n(M)$, 
\begin{enumerate}
\item  if $w_n(\theta) \geq 1 + \delta_n$ and $\| \theta- \theta_0\| \leq \lambda_n$
\begin{equation*}
E_n (\theta) = E_n  (\theta)\cap A_n(\theta)^c.
\end{equation*}
\item  if $w_n(\theta) \leq M_1$ for $M_1>0$, 
\begin{align*}
P_{\theta}(E_n' (\theta) \cap A_n (\theta)) \leq P_{\theta}(E_n (\theta) \cap A_n (\theta)) \leq P_{\theta}(E_n'' (\theta) \cap A_n (\theta))
\end{align*}
\end{enumerate}
\label{lemma1}
\end{lemma}

\begin{lemma}
Let     $M_1, M>0$, 
 we have  for $y \in \Omega_n(M)$, 
\begin{align*}
\sup_{\|\theta - \theta_0\|\leq M_1/ v_{n,k_0}} \frac{P_{\theta}(\tilde{E}_n (\theta))}{L_n} = O(1).
\end{align*}
\label{lemma2}
\end{lemma}

\begin{lemma}
For a given bandwidth $\epsilon_n,$ where $\epsilon_n<\lambda_n,$ the average probability of accepting in our accept/reject ABC algorithm, $\alpha_n$ is as follows.
\begin{align*}
\alpha_n = \int \pi(\theta) P_{\theta} \left( \norm{\eta(y)-\eta(z)} \leq \epsilon_n \right) d \theta \asymp L_n \epsilon_n^d
\end{align*}
\label{lemma3}
\end{lemma}

\subsection{Proof of Theorem \ref{theorem1}}

Let $M_0$ be an arbitrarily large constant, $\bar{\lambda}_n \geq M_0 \epsilon_n$ be a sequence going to 0 and such that $(\bar{\lambda}_n v_{n,k_0+1} )^{-\kappa}=o(L_n\epsilon_n^d) $, with $L_n$  defined in Assumption \ref{assump3} and $\kappa$ in Assumption \ref{assump4}. Note that such a $\bar \lambda_n$ exists since $ v_{n,k_0+1}^{-\kappa}=o(L_n\epsilon_n^d) $. 
Consider the event 
$$ \Omega_{n}(M) =\{ y; \|Z_{n}(\theta_0)\| \leq M \} , \quad Z_{n}(\theta_0) = D_n (\eta(y) - b(\theta_0)).$$ 
For all $\epsilon>0$, there exists $M_\epsilon>0$ such that $P_0(\Omega_{n}(M_\epsilon)^c)\leq \epsilon$. We fix $\epsilon$ and consider $M = M_\epsilon$. Hereafter we consider $y \in \Omega_n(M)$.  We wish to bound
\begin{align}
&\int_{\norm{b_{(2)}(\theta)-b_{(2)}(\theta_0)} \geq 2\bar{\lambda}_n} \pi_{\epsilon_n} \left( \theta | \eta(y) \right) d \theta = \frac{ \int_{\norm{b_{(2)}(\theta)-b_{(2)}(\theta_0)}\geq 2 \bar{\lambda}_n} \pi(\theta) P_{\theta}\left( \norm{\eta(y)-\eta(z)} \leq \epsilon_n \right) d \theta }{ \int \pi(\theta) P_{\theta}\left( \norm{\eta(y)-\eta(z)} \leq \epsilon_n \right) d \theta}.
\label{e18}
\end{align}
We first consider the numerator of \eqref{e18}. Decomposing, we can see that
\begin{align*}
\norm{\eta(y)-\eta(z)} \geq \norm{\eta_{(2)}(y)-\eta_{(2)}(z)} \geq \norm{b_{(2)}(\theta)-b_{(2)}(\theta_0)} - \left( \norm{\eta_{(2)}(y)-b_{(2)}(\theta_0)} + \norm{\eta_{(2)}(z)-b_{(2)}(\theta)} \right).
\end{align*}

Recall that $v_{n,j} \epsilon_n \rightarrow \infty $ for $j>k_0$,  and when $\theta$ belongs to  the set $\{\norm{b_{(2)}(\theta)-b_{(2)}(\theta_0)}\geq 2\bar{\lambda}_n \}$. Putting these together, we have on $\Omega_{n}(M)$:
\begin{align*}
\norm{\eta(y)-\eta(z)}\leq \epsilon_n \Rightarrow \left(  \norm{\eta_{(2)}(y)-b_{(2)}(\theta_0)} + \norm{\eta_{(2)}(z)-b_{(2)}(\theta)}  \right) \geq 2 \bar{\lambda}_n - \epsilon_n 
\end{align*}
which in turns implies that 
$$  \norm{\eta_{(2)}(y)-b_{(2)}(\theta_0)}  \geq 2\bar{\lambda}_n - 2\epsilon_n(1+o(1))  \geq 2 \bar{\lambda}_n \left(1-\frac{1}{M_0}\right)(1+o(1)).$$ 
Thus as soon as $M_0>2$
\begin{align}
&\int_{\norm{b_{(2)}(\theta)-b_{(2)}(\theta_0)}\geq 2 \bar{\lambda}_n} \pi(\theta) P_{\theta} \left( \norm{\eta(y)-\eta(z)} \leq \epsilon_n \right) d \theta \leq \int \pi(\theta) P_{\theta} \left( \norm{b_{(2)}(\theta)-\eta_{(2)}(z) } > \bar{\lambda}_n \right) d \theta \nonumber\\
&\leq (k-k_0)^\kappa\frac{ 1 }{ (\bar \lambda_n v_{n,k_0+1} )^\kappa } \int \pi(\theta) c(\theta) \pi(\theta) d\theta = o(L_n \epsilon_n^d),
\label{e19}
\end{align}
where the final inequality above comes from Assumption \ref{assump4}. 

To lower bound the denominator of \eqref{e18}, we simply apply Lemma \ref{lemma3}:
\begin{align}
&\int \pi(\theta) P_{\theta} \left( \norm{\eta(y)-\eta(z)} \leq \epsilon_n \right) d \theta \geq C L_n \epsilon_n^d.
\label{e20}
\end{align}
Going back to \eqref{e18}, applying \eqref{e19} and \eqref{e20} we find that
\begin{align*}
&\int_{\norm{b_{(2)}(\theta)-b_{(2)}(\theta_0)} \geq 2 \bar{\lambda}_n} \pi_{\epsilon_n} \left( \theta | \eta(y) \right) d \theta = o(1).\\
\end{align*}

By Assumption \ref{assump2}, the transformation $b(\cdot)_{(2)}$ is injective, which implies that

\begin{align*}
\int_{\norm{\theta-\theta_0} \geq 2 a^{-\frac{1}{2}} \bar{\lambda}_n } \pi_{\epsilon_n} \left( \theta | \eta(y) \right) d \theta = o(1),
\end{align*}
where $a$ is defined to be the largest eigenvalue of the matrix $\nabla b_{(2)}(\theta_0)^T \nabla b_{(2)}(\theta_0)$.

Defining $\lambda_n$ to be $2 a^{-\frac{1}{2}} \bar{\lambda}_n$, we have our result.

\subsection{Proof of Theorem \ref{theorem2}} 

\begin{proof}

The ABC posterior, $\pi_{\epsilon_n}(\theta|\eta(y)),$ can be expressed as 
\begin{align*}
\pi_{\epsilon_n}(\theta | \eta(y)) &= \frac{\pi (\theta) P_{\theta} (E_n) }{ \int_{\mathbb{R}^d} \pi (\theta) P_{\theta}(E_n) d \theta}, \quad E_n= \{ \norm{\eta(z)- \eta(y)} \leq \epsilon_n \},
\end{align*}

We define  $m(y,\theta)=Z_{n,(1)}(\theta_0) + D_{n,(1)}\nabla b_{(1)}(\theta-\theta_0)$.  We then can define the quantity $h_n(\theta)$ to be
\begin{align*}
h_n(\theta)  := L_n \gamma(m(y,\theta)) \mathbbm{1}_{ \norm{\nabla b_{(2)}(\theta_0) (\theta - \theta_0)} \leq \epsilon_n } \left( 1 - \frac{ \norm{\nabla b_{(2)}(\theta_0) (\theta - \theta_0)}^2}{\epsilon_n ^2} \right)^{\frac{R}{2}},
\end{align*}
where $\gamma: \mathbb{R}^{k_0} \rightarrow \mathbb{R}^+$, $L_n$ and $R$ are defined in Assumption \ref{assump3}. Note that $\|m(y,\theta)\|\leq M + C v_{n,k_0} \epsilon_n \leq C'$ for some $C,C'>0$ on $\Omega_n(M) \cap \{ \norm{\nabla b_{(2)}(\theta_0) (\theta - \theta_0)}\leq \epsilon_n \}$. In particular since $v_{n,k_0}\epsilon_n =o(1)$, 
\begin{equation}
m(y,\theta) = Z_{n,(1)}(\theta_0)+o(1). \label{mytheta:slow}
\end{equation}

Moreover, by Theorem \ref{theorem1}, $\pi_{\epsilon_n} (\|\theta - \theta_0\|> \lambda_n | y ) =o_{P_0}(1)$ and it is enough to control   
\begin{align*}
\Delta_n = \int_{\norm{\theta-\theta_0}<\lambda_n} \left| \frac{P_{\theta}(E_n)}{\int_{\norm{\theta-\theta_0}<\lambda_n}P_{\theta}(E_n)d \theta}  - \frac{ h_n(\theta) }{ \int_{\norm{\theta-\theta_0}<\lambda_n} h_n(\theta) d \theta } \right| d \theta.
\end{align*}
To prove our theorem, it will thus be sufficient to prove that $\Delta_n=o(1).$ 

In order to facilitate our demonstration, we define quantities $w_n(\theta),$ $V_n(\theta),$ and $V_n'(\theta)$ as $w_n(\theta) = \frac{\norm{\nabla b(\theta_0)(\theta-\theta_0)}^2}{\epsilon_n^2}, V_n(\theta) = L_n \mathbbm{1}_{\{ w_n(\theta) \leq 1 \}}(1-w_n(\theta))^{\frac{R}{2}},
V_n'(\theta) = L_n \mathbbm{1}_{\{ w_n(\theta) + \delta_n  \leq 1 \}}(1-w_n(\theta)-\delta_n)^{\frac{R}{2}}$, where $L_n=o(1)$ is defined in Assumption \ref{assump3}, $R$ is the constant defined in Assumption \ref{assump3}, and where $\delta_n$ is an $o(1)$ sequence defined in Lemma \ref{lemma1}. The quantity
$h_n(\theta)$ may then be expressed more simply as $h_n(\theta)= V_n(\theta) \gamma(m(y,\theta))$.

We have,
\begin{align*}
&\Delta_n = \int_{\norm{\theta-\theta_0}<\lambda_n} \left| \frac{P_{\theta}(E_n)}{\int_{\norm{\theta-\theta_0}<\lambda_n}P_{\theta}(E_n)d \theta}  - \frac{ h_n(\theta) }{ \int_{\norm{\theta-\theta_0}<\lambda_n} h_n(\theta) d \theta } \right| d \theta\\
&\leq  \int_{\norm{\theta-\theta_0}<\lambda_n} P_{\theta}(E_n)\left|\frac{1}{\int_{\norm{\theta-\theta_0}<\lambda_n}P_{\theta}(E_n)d \theta}- \frac{ 1 }{ \int_{\norm{\theta-\theta_0}<\lambda_n} h_n(\theta) d \theta} \right|  + \frac{ | P_{\theta}(E_n) - h_n(\theta)| }{ \int_{\norm{\theta-\theta_0}<\lambda_n} h_n(\theta) d \theta }   d \theta\\
&\leq 2   \frac{ \int_{\norm{\theta-\theta_0}<\lambda_n}  | P_{\theta}(E_n) - h_n(\theta)|d\theta }{ \int_{\norm{\theta-\theta_0}<\lambda_n} h_n(\theta) d \theta } := \frac{2 N_n}{D_n},
\end{align*}

In order to show that $\Delta_n=o(1),$ we show that $N_n= \frac{1}{L_n}\int_{\norm{\theta-\theta_0}<\lambda_n}  | P_{\theta}(E_n) - h_n(\theta)|d\theta =o(\epsilon_n^d)$ and $D_n = \frac{1}{L_n} \int_{\norm{\theta-\theta_0}<\lambda_n} h_n(\theta) d \theta  \geq C \epsilon_n^d.$

Recall that $\delta_n=o(1)$ slowly.  We then split the integral over $\left\{|\theta-\theta_0|\leq \lambda_n\right\}$ into 3 integrals over $\left\{w_n(\theta) \leq 1 - \zeta\right\}$, $\left\{1 - \zeta\leq w_n(\theta) \leq 1 + \sqrt{\delta_n}\right\}$ and $\left\{w_n(\theta) \geq 1 + \sqrt{\delta_n}\right\}$, where $\zeta>0$ is a fixed but arbitrarily small constant. This  leads to
$N_n \leq N_1 + N_2 + N_3$ with
\begin{align}\label{N123}
N_1 & = \frac{1}{L_n}  \int_{w_n(\theta) \leq 1 - \zeta} | P_\theta (E_n)- h_n(\theta) | d\theta \nonumber \\
N_2 & = \frac{1}{L_n}  \int_{ 1 - \zeta\leq w_n(\theta) \leq 1 + \delta_n}  | P_\theta (E_n)- h_n(\theta) | d\theta  \nonumber \\
N_3 & = \frac{1}{L_n}  \int_{w_n(\theta) > 1 + \delta_n}  \1_{\|\theta-\theta_0\|\leq \lambda_n}P_\theta (E_n)d\theta,
\end{align}
where the reduced integrand of $N_3$ above comes from the fact that $h_n(\theta) =0 $ when $w_n(\theta) > 1$.

From Lemma \ref{lemma1}, $P_\theta(E_n) \leq P_\theta(\tilde E_n)$ and using Lemma \ref{lemma2}, $ P_\theta(\tilde E_n)\leq CL_n$ uniformly over $w_n(\theta)\leq 2$ . By definition of $h_n$ we also have that 
$h_n(\theta) \leq L_n \gamma(m(y;\theta))$ and when $w_n(\theta)\leq 2$  and on $\Omega_n(M)$, $\gamma(m(y;\theta) )\leq \sup_{\|m\|\leq 2M} \gamma(m)<\infty$ so that $h_n/L_n$ is uniformly bounded. Hence on $\Omega_n$, there exists $C>0$ such that (using the change of variable $u = \nabla b_2(\theta_0)(\theta - \theta_0)$ and using the polar coordinates of $u$)

\begin{align}\label{N2bound}
    N_2 &\leq C \int \mathbbm{1}_{ 1 - \zeta \leq w_n(\theta) \leq 1 + \delta_n}d\theta \lesssim 
C \int_{\epsilon_n(1-\zeta)^{1/2}}^{\epsilon_n(1+ {\delta_n})^{1/2}}r^{d-1} dr  \lesssim \epsilon_n^d\zeta.
\end{align} 

We now study $N_1$. 
Firstly, we make use of the inequalities of Lemma \ref{lemma1} to upper and lower bound the quantity $P_{\theta}(E_n).$ We have
\begin{align}
P_{\theta}(E_n) &= P_{\theta}(E_n \cap A_n) + P_{\theta}(E_n \cap A_n^c) \leq P_{\theta}(E_n'' \cap A_n) + P_{\theta}(\tilde{E}_n \cap A_n^c) \nonumber\\
&\leq P_{\theta}(E_n'') +P_{\theta}(\tilde{E}_n \cap A_n^c),
\label{e4}
\end{align}
and
\begin{align}
P_{\theta}(E_n) &\geq P_{\theta}(E_n \cap A_n)  \geq P_{\theta}(E_n' \cap A_n)\nonumber\\
&= P_{\theta}(E_n')- P_{\theta}(E_n' \cap A_n^c) \geq P_{\theta}(E_n')-P_{\theta}(\tilde{E_n}\cap A_n^c).
\label{e5}
\end{align}
Combining \eqref{e4} and \eqref{e5}, and using the triangle inequality, we find 
\begin{align*}
|P_{\theta}(E_n)-h_n(\theta)| \leq \max \left\{ |P_{\theta}(E_n')-h_n(\theta)|,|P_{\theta}(E_n'')-h_n(\theta)| \right\} + P_{\theta}(\tilde{E}_n \cap A_n^c).
\end{align*}
Without loss of generality we assume that 
$\max \left\{ |P_{\theta}(E_n')-h_n(\theta)|,|P_{\theta}(E_n'')-h_n(\theta)| \right\}= \left|P_{\theta}(E_n')-h_n(\theta)\right|.$
It then follows that
\begin{align*}
    N_1 &\leq \frac{1}{L_n}  \int_{w_n(\theta) \leq 1 - \zeta} | P_\theta (E_n')- h_n(\theta) | d\theta + \frac{1}{L_n}  \int_{w_n(\theta) \leq 1 - \zeta}  P_\theta (\tilde E_n\cap A_n^c)d\theta
    \end{align*}
  Now using \eqref{limitEn},
 \begin{align*}
     P_\theta (E_n') &= P_{\theta}\left( \sum_{j=1}^{k_0}\frac{ [Z_{n,j}(\theta) - m_j(y;\theta) + O(v_{n,k_0}\epsilon_n^2)]^2 }{ v_{n,j}^2 } \leq \epsilon_n^2(1 - w_n(\theta) - \delta_n) \right)
     \end{align*}
To be able to apply Assumption \ref{assump3}, we need to replace $(1 - w_n(\theta) - \delta_n) $ by a constant $t$. To do so we consider the slices $S_i = \{ \theta: t_i< 1-w_n(\theta)-\delta_n \leq t_{i+1} \}$, where 
$$t_1= \zeta, \quad t_{i+1} = (1+ \zeta)t_i \quad i \leq  T_{\zeta}-1, $$
and $T_{\zeta}$ is the smallest integer satisfying $\sqrt{\zeta}(1+\zeta)^{T_{\zeta}} \geq 1$. 
We note that the union of sets $\cup_{i=1}^{(T_{\zeta}-1)}S_i$ covers $w_n(\theta) \leq 1-\zeta$ and that the sequence $\{t_1,\ldots,t_{T_{\zeta}} \}$ has the following properties.
\begin{align}
\frac{t_{i+1}}{t_i} &= 1 + \zeta = 1\quad \forall i \in \{ 1,\ldots, (T_{\zeta}-1) \} \noindent \\
\frac{t_i}{t_{i+1}} & = 1 - \frac{\zeta}{\zeta +1}  = 1  \quad  \forall i \in \{1,\ldots,(T_{\zeta}-1) \} . \label{t2}
\end{align}
Then, using that for $\theta \in S_i$, 
\begin{align*}
     P_\theta (E_n') &\leq  P_{\theta}\left( \sum_{j=1}^{k_0}\frac{ [Z_{n,j}(\theta) - m_j(y;\theta) + O(v_{n,k_0}\epsilon_n^2)]^2 }{ v_{n,j}^2 } \leq \epsilon_n^2t_{i+1} \right)\\
  P_\theta (E_n')      & \geq  P_{\theta}\left( \sum_{j=1}^{k_0}\frac{ [Z_{n,j}(\theta) - m_j(y;\theta) + O(v_{n,k_0}\epsilon_n^2)]^2 }{ v_{n,j}^2 } \leq \epsilon_n^2t_{i} \right)
     \end{align*}
and that using assumption \ref{assump3}, we bound on $\Omega_n(M)$, uniformly over $w_n(\theta) \leq 1 -\zeta$, 
\begin{align*}
     P_\theta (E_n') \leq L_n t_{i+1}^{R/2}( \gamma(  m_j(y;\theta)) + O(v_{n,k_0}\epsilon_n^2) ) + o(1)  \\
      P_\theta (E_n') \geq L_n t_{i}^{R/2}( \gamma(  m_j(y;\theta)) + O(v_{n,k_0}\epsilon_n^2) ) + o(1)  
     \end{align*}
     Moreover $\gamma$ is uniformly continuous over any compact and since $\|m_j(y;\theta) + O(v_{n,k_0}\epsilon_n^2) \|\leq 2M$ for $M$ large enough, 
     \begin{align*}
     P_\theta (E_n') \leq L_n t_{i+1}^{R/2}( \gamma(  m_j(y;\theta)  ) + o(1) )\\
      P_\theta (E_n') \geq L_n t_{i}^{R/2}( \gamma(  m_j(y;\theta) ) + o(1)  )
     \end{align*}
     which in turns implies that uniformly over $w_n(\theta) \leq 1 -\zeta$,
       \begin{align*}
\frac{ | P_\theta (E_n')  - h_n(\theta) | }{ L_n } &\leq  |t_{i+1}^{R/2} - t_i^{R/2}| \gamma(  m_j(y;\theta)  ) + o(1) \leq R \zeta (\zeta^{\frac R2-1}+1) \sup_{\|m\|\leq 2M} \gamma(m) +o(1) \\
&\lesssim \zeta^{\frac R2\wedge 1}+o(1).
     \end{align*}
We thus have 
\begin{align*}
    N_1 
     & \leq  
     \frac{1}{L_n}  \int_{w_n(\theta) \leq 1 - \zeta}  P_\theta (\tilde E_n\cap A_n^c)d\theta +O(\zeta^{\frac R2\wedge 1}\epsilon_n^d) .
\end{align*}

We now study $ P_\theta (\tilde E_n\cap A_n^c)$. By definition and on $\Omega_n(M) \cap  A_n^c$ , 
$$
\| D_{n,(2)}^{-1}Z_{n,(2)}(\theta) \| \geq \frac{ \delta_n \epsilon_n }{4} - \frac{ M }{v_{n,k_0+1}} \geq \frac{ \delta_n \epsilon_n }{8}$$
as soon as $v_{n,k_0+1}\delta_n \epsilon_n$ goes to infinity. Hence 
\begin{align*}
 P_\theta (\tilde E_n\cap A_n^c) \leq P_\theta \left( \left\{ \| D_{n,(2)}^{-1}Z_{n,(2)}(\theta) \| \geq \frac{\delta_n \epsilon_n }{8} \right\} \cap\left\{ \|\eta(z)_{(1)} -\eta(y)_{(1)} \|\leq \epsilon_n \right\} \right) 
 \end{align*}
 Using the proof of Lemma \ref{lemma2}, when $w_n(\theta) \leq 1$, 
 $$\eta_{(1)}(z)-\eta_{(1)}(y) = D_{n,1}^{-1}[Z_{n,(1)}(\theta) - m(y;\theta) + O(v_{n,1}\epsilon_n^2) ]  = D_{n,1}^{-1}[Z_{n,(1)}(\theta) - m(y;\theta)] + o(\epsilon_n) $$
 Moreover $m(y;\theta) =Z_{n,(1)}(\theta_0) + O(\epsilon_n)$, 
 so that $\|\eta_{(1)}(z)-\eta_{(1)}(y)\|\leq \epsilon_n$ implies that $\| D_{n,1}^{-1}[Z_{n,(1)}(\theta) - Z_{n,(1)}(\theta_0)] \| \leq M_1 \epsilon_n$ for some $M_1>0$. 
We then have using $\|Z_{n,(1)}(\theta_0)\|\leq M$
\begin{align*}
 P_\theta (\tilde E_n\cap A_n^c) &\leq E_\theta\left( \1_{ \| D_{n,1}^{-1}[Z_{n,(1)}(\theta) - Z_{n,(1)}(\theta_0)] \| \leq M_1 \epsilon_n} P_\theta \left(  \| D_{n,(2)}^{-1}Z_{n,(2)}(\theta) \| \geq \frac{\delta_n \epsilon_n }{8} \left| Z_{n,(1)} \right. \right) \right)\\
 & \leq  \sum_{j=k_0+1}^kE_\theta\left( \1_{ \| D_{n,1}^{-1}[Z_{n,(1)}(\theta) - Z_{n,(1)}(\theta_0)] \| \leq M_1 \epsilon_n} P_\theta \left( |Z_{n,j}(\theta) | \geq \frac{v_{n,j}\delta_n \epsilon_n }{8(k-k_0)} \left| Z_{n,(1)} \right. \right) \right)\\
 &\leq (k-k_0) \bar \rho_n(v_{n,k_0+1} \delta_n \epsilon_n t_0)P_\theta\left(   \|D_{n,1}^{-1}[Z_{n,(1)}(\theta) - Z_{n,(1)}(\theta_0)] \| \leq M_1 \epsilon_n \right)= o(L_n)
 \end{align*}
uniformly in $\theta$,  where the last two bounds come from Assumption \ref{assump5} and  Lemma \ref{lemma2}. Finally this leads to 
$$ N_1 = o(\epsilon_n^d) .$$

We now control $N_3$. Recall that
$$N_3 = \frac{1}{L_n}  \int_{w_n(\theta) > 1 +\delta_n} \1_{\|\theta-\theta_0\|\leq \lambda_n} P_\theta (E_n)d\theta.$$
We have, from Lemma \ref{lemma1}, 
$$N_3 = \frac{1}{L_n}  \int_{w_n(\theta) > 1 +\delta_n} \1_{\|\theta-\theta_0\|\leq \lambda_n} P_\theta (E_n\cap A_n^c)d\theta \leq  \frac{1}{L_n}  \int_{w_n(\theta) > 1 +\delta_n} \1_{\|\theta-\theta_0\|\leq \lambda_n} P_\theta (\tilde E_n\cap A_n^c)d\theta.$$
We then bound, for $\|\theta - \theta_0\|\leq \lambda_n$ and $y \in \Omega_n(M)$,
\begin{align*}
P_\theta (\tilde E_n\cap A_n^c) &= E_{\theta}\left[ \1_{\tilde E_n}(Z_{n,(1)}) P_\theta\left(\left. \norm{\eta_{(2)}(z) - b_{(2)}(\theta)} > \frac{\delta_n \epsilon_n}{4} \sqrt{w_n(\theta)} \right| Z_{n,(1)} \right)\right] \\
&\leq \sum_{j=k_0+1}^k E_{\theta}\left[ \1_{\tilde E_n}(Z_{n,(1)}) P_\theta\left( \left. |Z_{n,j}\theta)| > v_{n,j}\frac{\delta_n \epsilon_n}{4} \sqrt{w_n(\theta)} \right| Z_{n,(1)} \right) \right]\\
& \leq  \sum_{j=k_0+1}^k \bar \rho\left(v_{n,j}\frac{\delta_n \epsilon_n}{4} \sqrt{w_n(\theta)} \right) P_{\theta}\left[\tilde E_n \right] 
\end{align*}
since  $v_{n,j}\delta_n \epsilon_n \sqrt{w_n(\theta)} \gtrsim v_{n,k_0+1} \delta_n \epsilon_n \rightarrow \infty$. This implies that there exists $m_1>0$ such that  
 \begin{align*}
 N_{3,1} &:= \frac{1}{L_n}  \int_{w_n(\theta) > 1 +\delta_n} \1_{\|\theta-\theta_0\|\leq M_1/v_{n,k_0}} P_\theta (E_n\cap A_n^c)d\theta \\
 &\lesssim \frac{  1 }{ (v_{n,k_0+1}\delta_n )^\kappa } \int_{m_1 \epsilon_n\leq \|\theta - \theta_0\|\leq M_1/v_{n,k_0}} \|\theta - \theta_0\|^{-\kappa} P_\theta(\tilde E_n) d\theta \\
 & = O( (L_n (v_{n,k_0+1}\delta_n )^{-\kappa}) \int_{m_1 \epsilon_n}^{M_1/v_{n,k_0}}r^{d-1-\kappa} dr \\
 &=O( L_n (v_{n,k_0+1}\delta_n\epsilon_n)^{-\kappa} \epsilon_n^d ) = o( L_n \epsilon_n^d ) 
\end{align*}
since  $\kappa >d$. 
We also have 
\begin{align*}
 N_{3,2} &:= \frac{1}{L_n}  \int_{\lambda_n \geq\|\theta-\theta_0\|\leq M_1/v_{n,k_0}}  P_\theta (E_n\cap A_n^c)d\theta \\
 &\lesssim \frac{  1 }{ (v_{n,k_0+1}\delta_n)^\kappa } \int_{M_1/v_{n,k_0} \leq \|\theta - \theta_0\|\leq \lambda_n} \|\theta - \theta_0\|^{-\kappa} P_\theta(\tilde E_n) d\theta \\
 & = O(  (v_{n,k_0+1}\delta_n )^{-\kappa}) \int_{M_1/v_{n,k_0}}^{\lambda_n} r^{d-1-\kappa} dr \\
 &\lesssim  \frac{  v_{n,k_0}^{-d+\kappa} }{ (v_{n,k_0+1}\delta_n)^{\kappa} } \lesssim
 \epsilon_n^d \delta_n^{-\kappa} \left(\frac{  v_{n,k_0} }{ v_{n,k_0+1} }\right)^{\kappa -d} (\epsilon_n v_{n,k_0+1})^{-d} =o(L_n \epsilon_n^d )
\end{align*}
by assumption on $v_{n,k_0}, v_{n,k_0+1}, \epsilon_n$.

We now consider the order of $D_n.$
\begin{align}
&D_n := \frac{1}{L_n} \int_{\norm{\theta-\theta_0}<\lambda_n} h_n(\theta) d \theta \nonumber\\
&= \int_{\norm{\theta-\theta_0}<\lambda_n} \mathbbm{1}_{\{ w_n(\theta) \leq 1 \}}(1-w_n(\theta))^{\frac{R}{2}} \gamma \left(\eta_{(1)}(y)-b_{(1)}(\theta_0) \right) d \theta \nonumber\\
&\geq C \int_{\norm{\theta-\theta_0}<\lambda_n} \mathbbm{1}_{\{ w_n(\theta) \leq 1 \}}(1-w_n(\theta))^{\frac{R}{2}} d \theta \nonumber\\
& \geq C \int_{\frac{1}{2} \leq w_n(\theta)\leq 1} \frac{1}{2} d \theta \nonumber\\
&=(C+o(1))\epsilon_n^d.
\label{e25}
\end{align}
The third line of the set of equations above comes from the fact that $\gamma(\cdot)$ is lower bounded by a positive constant. 

Combining the upper bound on $N_n$  and \eqref{e25}, we have
\begin{align}
\Delta_n = \frac{N_n}{D_n} = \frac{o(\epsilon_n^d)}{C(\epsilon_n^d)} = o(1).
\label{Delta}
\end{align}

We thus have that the ABC posterior, $\pi_{\epsilon_n}(\theta)$, converges in distribution to
\begin{align*}
\frac{h_n(\theta)}{\int_{\norm{\theta-\theta_0}<\lambda_n}h_n(\theta)d\theta} \propto \bm{1}_{\{ \norm{ \nabla b_{(2)}(\theta_0) (\theta - \theta_0)} \leq \epsilon_n \}} \left( 1 - \frac{ \norm{\nabla b_{(2)}(\theta_0)(\theta-\theta_0)}^2}{\epsilon_n^2} \right)^{\frac{R}{2}}
\end{align*}
as wanted.

\end{proof}

\subsection{Proof of Theorem \ref{th:LS}} \label{pr:th:LS}

\begin{proof}

Let $m_0 = E_{\theta_0}(Z_n(\theta_0))$. Minimizing in $B, \beta_0$ , $ L(\beta, \beta_0)$ is equivalent to minimizing in $B $
\begin{equation*} 
 E_n[\|\theta-E_n(\theta) - B \tilde S \|^2] = \sum_{j=1}^d
 E_n[(\tilde\theta_j- \beta(j)^T\tilde S]^2] ,
\end{equation*}
which we  write $\sum_{j=1}^dL_j(\beta(j))$ and where $\beta(j)$ is the $j$-th row of $B$, $\tilde \theta = \theta -E_n(\theta)
$ and $\tilde S = S - E_n(S)$.  We can thus study the terms $L_j$ separately. Let $j\leq d$, we have 
\begin{equation*} 
 L_j(\beta(j)) =
 V_n(\theta_j) +  \beta(j)^T E_n(\tilde S \tilde S^T ) \beta(j) - 2 \beta(j)^T E_n( \tilde S \tilde \theta_j ) 
 \end{equation*}
%and writing $u =( \theta - E_n(\theta))/\epsilon_n$, $ R(u) = b(\theta(u)) - b_0 - \epsilon_n\nabla b_0 (\theta - \theta_0) =  b(\theta(u)) - b_0 - \epsilon_n\nabla b_0 [u + E_n(\theta-\theta_0)
 \begin{equation} \label{tildeS}
\tilde S =  \epsilon_n(  \nabla b(\theta_0) u + \epsilon_n R(u) ) + D_n^{-1} [Z_n(\theta) - E_n(Z_\theta)  ]
\end{equation}
where $\epsilon_n^2 R(u)  = b(\theta(u)) - E_n(b(\theta(u)) - \epsilon_n\nabla b(\theta_0) u  = O( \epsilon_n^2 \|u\|^2)$. 
We first study $E_n( \tilde S \tilde S^T )$. First note that 
\begin{equation}\label{EZ}
 E_n(Z_\theta)= E_n ( E_\theta(Z_n(\theta) ) = E_{\theta_0}(Z_n(\theta_0) + O( \|\theta-\theta_0 \|) = m_0+ O( \|\theta-\theta_0 \|). 
\end{equation}

%\begin{equation}\label{EZ}
%= E_n ( E_\theta(Z_n(\theta) ) = E_{\theta_0}(Z_n(\theta_0) + O( \|\theta-\theta_0 \|) = m_0+ O( \|\theta-\theta_0 \|). 
%\end{equation}
Then writing $\tilde Z_n(\theta) = Z_n(\theta) - E_n(Z_n(\theta))$, 
\begin{equation*}
\begin{split}
E_n( \tilde S \tilde S^T )& = \epsilon_n^2 \nabla b(\theta_0) E_n( u u^T ) \nabla b(\theta_0)^T + D_n^{-1}\tilde Z_n(\theta)\tilde Z_n(\theta)^T D_n^{-1}  +2 \epsilon_n D_n^{-1}E_n(\tilde Z_n(\theta) u^T)\nabla b(\theta_0)^T \\
&\qquad+  2 \epsilon_n^2 D_n^{-1}E_n(\tilde Z_n(\theta)R(u)^T)+O(\epsilon_n^3 E_n(\|u\|\|R(u)\|) 
\end{split}
\end{equation*}
Since for any function $H(u)$, using $\tilde Z_n(\theta) = Z_n(\theta) - E_\theta(Z_n(\theta)) + E_\theta(Z_n(\theta))  - E_n(Z_n(\theta))$ together with \eqref{EZ}
$$E_n(\tilde Z_n(\theta)H(u)) = E_n( [E_\theta(  Z_n(\theta))-E_n(Z_n(\theta))] H(u) )= O(\epsilon_n E_n(\|u\||H(u)|),$$
we obtain that 
\begin{equation}\label{SS}
\begin{split}
E_n( \tilde S \tilde S^T )& = \epsilon_n^2 \nabla b(\theta_0) E_n( u u^T ) \nabla b(\theta_0)^T + D_n^{-1}\tilde Z_n(\theta)\tilde Z_n(\theta)^T D_n^{-1} \\
 & \quad + O\left(\epsilon_n^2\frac{  E_n(\|u\|^3|)  }{ v_{n,1} } \right)
\end{split}
\end{equation}

We write $\tilde S_{(1)} = (\tilde S_1,\ldots, \tilde S_{k_0})$ and $\tilde S_{(2)} = \tilde S_{k_0+1}, \ldots, \tilde S_J$. 
\begin{equation*}
\begin{split}
E_n( \tilde S_{(2)} \tilde S_{(2)}^T )& = \epsilon_n^2 \nabla b_2(\theta_0) E_n( u u^T ) \nabla b_2(\theta_0)^T + O\left( \frac{ \epsilon_n^2 }{ v_{n,1} }  + \frac{ 1 }{ v_{n,k_0+1}^2 } \right) =  \epsilon_n^2 \nabla b_2(\theta_0) E_n( u u^T ) \nabla b_2(\theta_0)^T+o(\epsilon_n^2), 
\end{split}
\end{equation*}
and
\begin{align*}
 E_n [ \tilde S_{(1)} \tilde S_{(1)}^T ] &=D_{n,1}^{-1}C_{1,1} D_{n,1}^{-1} + \epsilon_n^2  \nabla b_1(\theta_0)E_n(u u^T)\nabla b_1(\theta_0)^T + O(\epsilon_n^3) + O\left( \frac{ \epsilon_n^2 }{ v_{n,1}  } \right)\\
 \end{align*}
\begin{equation*}
\begin{split}
E_n( \tilde S_{(2)} \tilde S_{(1)}^T )& = \epsilon_n^2 \nabla b_2(\theta_0) E_n( u u^T ) \nabla b_1(\theta_0)^T + D_{n,2}^{-1}C_{2,1} D_{n,1}^{-1}
+ O\left( \frac{ \epsilon_n^2 }{ v_{n,1} } \right)
\end{split}
\end{equation*}
Also 
\begin{equation}\label{SS}
\begin{split}
E_n( \tilde S \tilde \theta_j )& = \epsilon_n^2 \nabla b(\theta_0) E_n( u u_j ) + \epsilon_nD_n^{-1}E_n(\tilde Z_n(\theta)u_j) + O\left(\epsilon_n^3 \right) \\
 & =\epsilon_n^2 \nabla b(\theta_0) E_n( u u_j ) + \epsilon_nD_n^{-1} O( \epsilon_n) + O\left(\epsilon_n^3 \right)
\end{split}
\end{equation}
Finally we obtain that 
\begin{align*}
L_j(\beta(j)) & =V_n(\theta_j) +  \beta_{(1)}(j)^T[D_{n,1}^{-1}C_{1,1} D_{n,1}^{-1} + \epsilon_n^2  \nabla b_1(\theta_0)E_n(u u^T)\nabla b_1(\theta_0)^T+ O( \epsilon_n^2/v_{n,1})] \beta_{(1)}(j) \\
 &\qquad +\epsilon_n^2  \beta_{(2)}(j)^T \nabla b_2(\theta_0)E_n(u u^T) \nabla b_2(\theta_0)^T \beta_{(2)}(j) + 2 \beta_{(1)}(j)^TD_{n,1}^{-1}C_{1,2} D_{n,2}^{-1}\beta_{(2)}(j)\\
 &\qquad  - 2 \epsilon_n^2 \beta(j)^T \nabla b(\theta_0) E_n( u u_j ) + O\left(\epsilon_n^3 \right)+ O( \epsilon_n^2 / v_{n,1} )
\end{align*} 
with $C = E_n( \tilde Z_n(\theta) \tilde Z_n(\theta)^T ) $, $C_{1,1}$ is the top left submatrix of dimension $k_0$, $C_{2,2}$ the bottom right with dimension $k-k_0$ and $C_{1,2}$ the top right with dimensions $k_0, k-k_0$. 
Note that 
$  \nabla b_2(\theta_0)E_n(u u^T) \nabla b_2(\theta_0) =  \nabla b_2(\theta_0)[E_h(u u^T) +o(1)]\nabla b_2(\theta_0)$,  where 
$$E_h(uu^T) = \frac{ \int_{B_2} uu^T(1 - \|\nabla b_2(\theta_0) u\|^2 )^{R/2} du  }{  \int_{B_2}(1 - \|\nabla b_2(\theta_0) u\|^2 )^{R/2} du} , \quad B_2 = \{ u\in \mathbb R^d; \, \|\nabla b_2(\theta_0) u\|\leq 1 \},$$
and is therefore positive semi-definite.

Minimizing $L_j(\beta(j))$ boils down to minimizing  in $ \tilde \beta_2 = \beta_{(2)}(j), \tilde \beta_1 = D_{n,1}^{-1}\beta_{(1)}(j)/\epsilon_n$
\begin{align*}
 \tilde L(\tilde \beta)  &= \tilde \beta_1^T[C_{1,1}+  D_{n,1}^{-1} \nabla b_1(\theta_0)E_n(u u^T)\nabla b_1(\theta_0)^TD_{n,1}^{-1}  + O( 1/v_{n,1}^3)] \tilde\beta_1 \\
 & +  \beta_2^T \nabla b_2(\theta_0)E_n(u u^T) \nabla b_2(\theta_0)^T \beta_2 + 2\epsilon_n^{-1} \beta_1 C_{1,2} D_{n,2}^{-1}\beta_2  - 2 \beta_2^T \nabla b_2(\theta_0)E_n( u u_j) - 2 \epsilon_n \beta_1^T D_{n,1}  \nabla b_2(\theta_0)E_n( u u_j)  + O( 1 / v_{n,1} )\\
  & = \tilde \beta_1^TC_{1,1}\tilde \beta_1^T + \beta_2^T \nabla b(\theta_0)E_n(u u^T) \nabla b_2(\theta_0)^T \beta_2 - 2 \beta_2^T \nabla b_2(\theta_0)E_n( u u_j) + O(1/v_{n,1})
 \end{align*} 
Any minimum verifies 
$$ C_{11} \tilde \beta_1 = O(1/v_{n,1}), \quad \nabla b_2(\theta_0)^T \beta_2 = E_n(u u^T)^{-1} E_n( u u_j) + O(1/v_{n,1})$$ 
In particular the minimum with smaller norm satisfies at the limit 
$$  \beta_1^* = 0, \quad \nabla b_2(\theta_0)^T \beta_2^* = E_n(u u^T)^{-1} E_n( u u_j)= e_j $$
which is the $j$-th vector in the canonical bases of $\mathbb R^d$. This proves the first part of Theorem \ref{th:LS}. We now study 
$\theta' - \theta_0 = \theta - \theta_0 - B^*(S-S_0)$ .

We have for all $j\leq d$
\begin{equation*}
\begin{split}
\theta_j' - \theta_{0j}  &= \theta_j - \theta_{0j}  - (\beta_2^*(j))^T \nabla b_2( \theta_0) (\theta-\theta_0) + \beta_2^*(j)^TD_{n,2}^{-1} (Z_{n,(2)}(\theta) - Z_{n,(2)}(\theta_0)) + O( \epsilon_n^2 ) \\
& =\beta_2^*(j)^TD_{n,2}^{-1} (Z_{n,(2)}(\theta) - Z_{n,(2)}(\theta_0)) + O( \epsilon_n^2 ) 
\end{split} 
\end{equation*}

\end{proof}

\subsection{Proof of Lemma \ref{lemma1}}

\begin{proof}
Throughout the proof $C$ denotes a generic constant whose value is of no importance and can vary from one line to the next. 

Let $ \delta_n=o(1)$ such that $ \delta_n v_{n,k_0+1}\epsilon_n\rightarrow \infty$.  Let $M>0$ and consider $y \in \Omega_n(M)$, then for all $\|\theta - \theta_0\|\leq \lambda_n = o(1)$, 
\begin{align*}
 \norm{ \eta_{(2)}(z)-\eta_{(2)}(y)}^2 &= \norm{ \nabla_2(\theta_0)(\theta - \theta_0)(1+O(\lambda_n)) ) + D_{n,(2)}^{-1}( Z_{n,(2)}(\theta)-Z_{n,(2)}(\theta_0)}^2 \\
  & \geq  \epsilon_n^2 \left( w_n(\theta) (1- C \lambda_n)+ \norm{ \frac{ D_{n,(2)}^{-1}( Z_{n,(2)}(\theta)-Z_{n,(2)}(\theta_0))}{ \epsilon_n} }^2\right. \\
&\qquad \left. 
  - 2  \sqrt{ w_n(\theta)(1 -C\lambda_n) } \norm{ \frac{ D_{n,(2)}^{-1}( Z_{n,(2)}(\theta)-Z_{n,(2)}(\theta_0)}{ \epsilon_n} } \right) 
 \end{align*}
 Hence if $w_n(\theta) \geq 1+\delta_n$, on  $A_n $,$$\norm{ \frac{ D_{n,(2)}^{-1} Z_{n,(2)}(\theta)}{ \epsilon_n} } \leq  \delta_n\sqrt{ w_n(\theta)}/4$$
 so that 
 \begin{equation}
 \begin{split}
 \norm{ \eta_{(2)}(z)-\eta_{(2)}(y)}^2
  & \geq   \epsilon_n^2w_n(\theta) \left( 1 -  C\lambda_n-   \delta_n \sqrt{1 + C\lambda_n}/2   \right) \\
  &\geq \epsilon_n^2 ( 1+ \delta_n)(1 -  C\lambda_n-   \delta_n \sqrt{1 + C\lambda_n}/2 ) \\
  &> \epsilon_n^2 
\label{e29}
\end{split}
\end{equation}
if $w_n(\theta) > 1 + \delta_n$ and as soon as $C\lambda_n < \frac{\delta_n}{3}$ and $\delta_n$ is small enough. 
Hence part 1 of Lemma \ref{lemma1} is proved. 

We now prove part 2. Let $\theta$ be such that $w_n(\theta)\leq M_1$. We omit $\theta$ in the notations $E_n, E_n', A_n, \tilde E_n $. Using the same computations as above , on $A_n$, if $w_n(\theta)> 1$, 
 \begin{align*}
 \norm{ \eta_{(2)}(z)-\eta_{(2)}(y)}^2
  & \geq   \epsilon_n^2w_n(\theta) \left( 1 -  C\lambda_n-   \delta_n \sqrt{1 + C\lambda_n}/2   \right) \geq \epsilon_n^2w_n(\theta) (1 - \delta_n) 
\end{align*}
and similarly 
 \begin{align*}
 \norm{ \eta_{(2)}(z)-\eta_{(2)}(y)}^2
  & \leq   \epsilon_n^2w_n(\theta) ( 1 + C\lambda_n)(1  +   \delta_n /4)^2    \leq \epsilon_n^2w_n(\theta) (1 + \delta_n) .
\end{align*}
Also if $ \delta_n \leq w_n(\theta) \leq 1$, 
\begin{align*}
 \norm{ \eta_{(2)}(z)-\eta_{(2)}(y)}^2
  & \geq  \epsilon_n^2 \left( \sqrt{w_n(\theta) (1- C \lambda_n) } - \delta_n/4 \right)^2 \\
   \norm{ \eta_{(2)}(z)-\eta_{(2)}(y)}^2
  & \leq  \epsilon_n^2 \left( \sqrt{w_n(\theta) (1+ C \lambda_n) } + \delta_n/4 \right)^2 
 \end{align*}
 so that if,
 $\norm{ \eta_{(1)}(z)-\eta_{(1)}(y)}^2 \leq \epsilon_n^2 ( 1 - w_n(\theta) - \delta_n)$, $w_n(\theta)\leq 1$ and 
 $$\norm{\eta(z)-\eta(y)}^2 \leq \epsilon_n^2[ C \lambda_n +\sqrt{1+ C \lambda_n } \delta_n/2 - \delta_n \leq \epsilon_n^2$$
 by choosing $\lambda_n \leq c\delta_n$ with $c$ small enough. 
 Hence $E_n' \cap A_n \subset E_n \cap A_n$. Similar arguments imply that $E_n \cap A_n \subset E_n" \cap A_n$.
 
\end{proof}

\subsection{Proof of Lemma \ref{lemma2} }
\begin{proof}
We have
$$\eta_{(1)}(z)-\eta_{(1)}(y) = D_{n,1}^{-1}[Z_{n,(1)}(\theta) - Z_{n,(1)}(\theta_0)] + b_{(1)}(\theta) - b_{(1)}(\theta_0)$$ 
so that if $\| \theta - \theta_0\| \leq v_{n,k_0}^{-1}M_1\wedge \delta_n$, 
$$\eta_{(1)}(z)-\eta_{(1)}(y) = D_{n,1}^{-1}[Z_{n,(1)}(\theta) - m(y;\theta) + O(v_{n,k_0}^{-1}\wedge v_{n,k_0}\delta_n^{2}) ] =   D_{n,1}^{-1}[Z_{n,(1)}(\theta) - m(y;\theta) + o(1) ]$$
where 
$ m(y;\theta) =Z_{n,(1)}(\theta_0) + D_{n,1}\nabla b_1(\theta_0)(\theta - \theta_0)$. and on $\Omega_n(M)$ 
$ \| m(y;\theta) +o(1)  \| \leq 2M$ by choosing $M$ large enough. 
\begin{align*}
 P_{\theta}(\tilde{E}_n) 
 =  P_{\theta}\left( \sum_{j=1}^{k_0}\frac{ [Z_{n,j}(\theta) - m_j(y;\theta) + o(1)]^2 }{ v_{n,j}^2 } \leq \epsilon_n^2 \right)
\end{align*}
It implies in particular that with $K$ the ball in $\mathbb R^{k_0}$ centered at 0 and with radius $2M$, uniformly over $\|\theta-\theta_0\|\leq \lambda_n$, 
\begin{align} \label{limitEn}
\left| \frac{  P_{\theta}(\tilde{E}_n) }{ L_n } - \gamma( m_j(y;\theta)+o(1))  \right| 
& \leq \sup_{m \in K} \left| \frac{P_{\theta}\left( \sum_{j=1}^{k_0}\frac{ [Z_{n,j}(\theta) - m_j ]^2}{ v_{n,j}^2 }\leq \epsilon_n^2 \right) }{L_n} - \gamma (m) \right| \\
&=o(1),
\end{align}
where the last equality comes  from Assumption \ref{assump3}.

It follows in particular that 
\begin{align*}
&\sup_{\|\theta- \theta_0\|\leq \lambda_n} \norm{\frac{P_{\theta}(\tilde{E}_n)}{L_n}} \leq \sup_{m\in K}\gamma(m) + o(1)=O(1).
\end{align*}

\end{proof}

\subsection{Proof of Lemma \ref{lemma3}}
\begin{proof}
We have
\begin{align*}
\alpha_n & = \int \pi(\theta) P_{\theta} \left( \norm{ \eta(z)-\eta(y) } < \epsilon_n \right) d\theta\\
& = \int_{\norm{\theta - \theta_0} < \lambda_n } h_n(\theta) d \theta + o \left( L_n \epsilon_n^d \right)\\
& = L_n \gamma \left( \eta_{(1)}(y)-b_{(1)}(\theta_0) \right) \int_{\norm{\theta - \theta_0} < \lambda_n ; w_n(\theta) \leq 1} (1 - w_n(\theta))^{\frac{R}{2}} d \theta + o \left( L_n \epsilon_n^d \right) \\
& = L_n \epsilon_n^d \gamma \left( \eta_{(1)}(y)-b_{(1)}(\theta_0) \right) \det \left( \nabla b_{(2)}(\theta_0)^T \nabla b_{(2)}(\theta_0) \right) \int_{u \leq 1} (1-u^2)^{\frac{R}{2}} d u + o\left( L_n \epsilon_n^d \right)\\
&= L_n \epsilon_n^d \gamma \left( \eta_{(1)}(y)-b_{(1)}(\theta_0) \right) \frac{1}{2} \det \left( \left( \nabla b_{(2)}(\theta_0) \nabla b_{(2)}(\theta_0) \right)^{\frac{1}{2}} \right) \text{Beta} \left( \frac{1}{2},\frac{R}{2}+1 \right) + o \left( L_n \epsilon_n^d \right)\\
&= ( C+o(1))L_n \epsilon_n^d .
\end{align*}

The third line of the set of equations above comes from \eqref{Delta}. The fourth line comes from the definition of $h_n(\theta).$ The fifth line comes from a change of variables.
\end{proof}

\end{document}